\documentclass[aps,pra,longbibliography,twocolumn,showpacs]{revtex4-1}

\usepackage[latin1]{inputenc}
\usepackage[pdftex]{graphicx}
\usepackage{bbold}
\usepackage{color}
\usepackage{amsmath}
\usepackage{amssymb}
\usepackage{amsfonts}

\newcommand{\ket}[1]{\vert{#1}\rangle}
\newcommand{\bra}[1]{\langle{#1}\vert}
\newcommand{\bv}[1]{\boldsymbol{#1}}
\newcommand{\up}{\uparrow}
\newcommand{\dn}{\downarrow}

\newcommand{\gen}{\mathcal{L}}
\newcommand{\JJ}{\mathcal{J}}
\newcommand{\Dr}{\mathcal{D}_{\rm r}}

\newcommand{\rr}{\boldsymbol{r}}

\newcommand{\sv}{\boldsymbol{s}}

\newcommand{\rmd}{\mathrm{d}}
\newcommand{\rme}{\mathrm{e}}
\newcommand{\rmi}{\mathrm{i}}

\newcommand{\avt}[1]{\langle #1\rangle_{\rm tr}}
\newcommand{\avp}[1]{\langle #1\rangle_{\rm p}}

\definecolor{darkred}{rgb}{0.8,0.1,0.1}

\begin{document}

\title{Sensing of single nuclear spins in random thermal motion with proximate nitrogen-vacancy centers}

\author{M.~Bruderer, P.~Fern\'{a}ndez-Acebal, R.~Aurich and M.~B.~Plenio}
\address{Institut für Theoretische Physik, Albert-Einstein Allee 11, Universität Ulm, 89069 Ulm, Germany}


\begin{abstract}
Nitrogen-vacancy (NV) centers in diamond have emerged as valuable tools for sensing and polarizing spins.
Motivated by potential applications in chemistry, biology and medicine we show that NV-based sensors
are capable of detecting single spin targets even if they undergo diffusive motion in an ambient thermal
environment. Focusing on experimentally relevant diffusion regimes we derive an effective model for the
NV-target interaction, where parameters entering the model are obtained from numerical simulations of the
target motion. The practicality of our approach is demonstrated by analyzing two realistic experimental
scenarios: (i)~time-resolved sensing of a fluorine nuclear spin bound to an N-heterocyclic carbene-ruthenium
(NHC-Ru) catalyst that is immobilized
on the diamond surface and (ii)~detection of an electron spin label by an NV center in a nanodiamond, both
attached to a vibrating chemokine receptor in thermal motion. We find in particular that the detachment of
a fluorine target from the NHC-Ru carrier molecule can be monitored with a time resolution of a few seconds.
\end{abstract}

\maketitle

%%%%%%%%%%%%%%%%%%%%%%%%%%%%%%%%%%%%%%%%%%%%%%%%%%%%%%%%%%%%%%%%%%%%%%
%%%%%%%%%%%%%%%%%%%%%%%%%%%%%%%%%%%%%%%%%%%%%%%%%%%%%%%%%%%%%%%%%%%%%%

\section{Introduction}

Single-molecule observation techniques have provided many important insights into the details of fundamental
chemical and biological processes. Specifically, the investigation of single-molecule catalytic
reactions in various environments has been made possible by several, primarily spectroscopic methods. To give
some examples, fluorescence microscopy has been used to observe single-molecule enzymatic dynamics~\cite{lu1998single,
flomenbom2005stretched,velonia2005single} and to determine the spatial distribution of catalytic activity on
crystal surfaces~\cite{roeffaers2006spatially}. Moreover, scanning tunneling microscopy has demonstrated its
potential for monitoring oxidation catalysis in real time at liquid-solid interfaces~\cite{hulsken2007real}.

\begin{figure}[b]
\centering
\includegraphics[width=240pt]{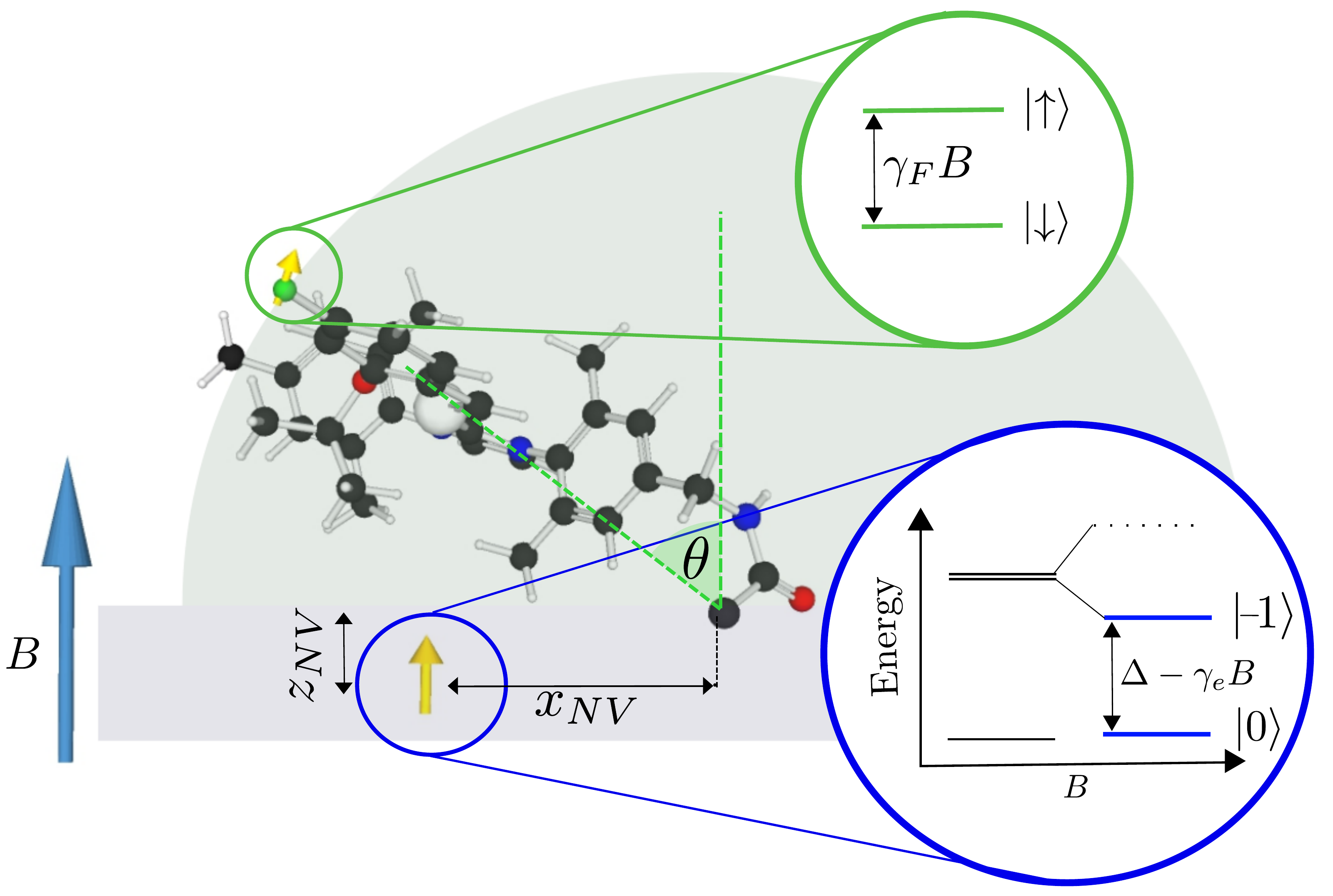}
\caption{Sensing of a single fluorine atom undergoing random thermal motion using a single near-surface
nitrogen-vacancy center in bulk diamond: The fluorine-containing molecule (N-heterocyclic carbene-ruthenium complex)
is covalently attached to the diamond and rotates freely around its anchor point, confining the thermal
motion of the fluorine to a spherical surface. The nuclear spin of the fluorine and the microwave-driven
electron spin of the NV center interact resonantly at the Hartmann-Hahn matching condition. Time-resolved
sensing of the fluorine in thermal motion is achieved by observing the polarization transfer
between the spins.}
\label{ModelSketch}
\end{figure}

More recently, nitrogen-vacancy (NV) centers in bulk diamond~\cite{jelezko2006single} have been used for
spin resonance spectroscopy with a sensitivity ultimately reaching the single-molecule level~\cite{staudacher2013nuclear,
muller2014nuclear,shi2015single}, which makes them a promising new tool for probing the dynamics of chemical
processes. This is particularly true for single-site catalytic reactions~\cite{basset2009modern}, in which case
the catalyst is immobilized on a surface while maintaining its full catalytic functionality. The probing of catalysis
on a diamond surface can be accomplished, in principle, by using surface-implanted NV centers to monitor single
nuclear spins of the catalyst (see Fig.~\ref{ModelSketch}), which yields valuable information about the catalytic
reaction, e.g.,~about conformational changes. As the probing method is label-free it is also suitable for more
general biosensing applications~\cite{schirhagl2014nitrogen,wu2015}. While this prospect is certainly appealing, the
important question remains of whether NV-based sensing of nuclear spins is still efficiently achievable under ambient
thermal conditions, i.e., when the nuclear target spins exhibit random thermal motion.

%%%%%%%%%%%%%%%%%%%%%%%%%%%%%%%%%%%%%%%%%%%%%%%%%%%%%%%%%%%%%%%%%%%%%%
%%%%%%%%%%%%%%%%%%%%%%%%%%%%%%%%%%%%%%%%%%%%%%%%%%%%%%%%%%%%%%%%%%%%%%

In this paper we show that the detection of nuclear spins (and target spins in general) undergoing thermal
diffusion is indeed feasible under realistic experimental conditions. We focus our attention on the nuclear
spin of a fluorine atom covalently bound to a ruthenium-based catalyst (NHC-Ru complex~\cite{dragutan2007nhc})
and analyze the time-resolved sensing of a possible detachment of the fluorine from the catalyst.
%%%
Our theoretical study is strongly motivated by recent experimental progress: Decoupling schemes for the
electron spin of the NV now offer the possibility to resonantly amplify interactions between the NV center
and selected spin targets and to suppress decoherence caused by surrounding spins. In this way, single nuclear
spins located in the diamond lattice are routinely detected by employing either pulsed or
continuous dynamical decoupling~\cite{kolkowitz2012sensing,taminiau2012detection,london2013detecting}.
The achieved sensitivity of NV centers has moreover been shown to be sufficient to discern even single spin
targets or molecules on the diamond surface~\cite{staudacher2013nuclear,muller2014nuclear,shi2015single}
and also to detect spin ensembles surrounding NV centers hosted in nanodiamonds~\cite{mcguinness2013ambient}.

To make the analysis more definite we consider continuous dynamical decoupling of the NV center in
combination with the Hartmann-Hahn double resonance (HHDR) scheme~\cite{hartmann1962nuclear}, which has
been successfully used for detecting single static target spins~\cite{london2013detecting}.
At first we substantially extend the existing theory on the HHDR scheme for NV centers~\cite{cai2013diamond,london2013detecting}
by including the effects caused by the motion of the target spin. Making use of a stochastic description of the
diffusive target we derive an effective quantum master equation for the evolution of the spin system formed
by the NV center and the target. We find in particular that the evolution of the spin system for a target
undergoing fast diffusion is remarkably similar to the case of a static target, but with an effective
\emph{averaged} NV-target coupling replacing the static NV-target coupling. Details of the motion of the target are
obtained from  anisotropic network model~(ANM) simulations~\cite{tirion1996large,doruker2000dynamics,atilgan2001anisotropy}
of the entire NHC-Ru complex. The results of the ANM simulation allow us to generate stochastic trajectories of the target
spin only, which in turn can be used for estimating the parameters of the effective quantum master equation.

When applied to the fluorine attached to the NHC-Ru complex our approach yields that the effective NV-target
coupling can reach values up to a few hundred kilohertz and therefore is sufficiently strong for observable
HHDR polarization transfers between the spins to be achieved. The motion of the target spin also results in
diffusion-induced decoherence of the NV spin, which is however negligible. As a result, the detachment
of a target spin from the NHC-Ru carrier molecule can be monitored with a time resolution of a few seconds.
In addition, we analyze an alternative scenario, where an electron spin label is attached to a CXCR4
chemokine receptor, which plays an important role in infection processes~\cite{wu2010structures}.
Unlike in the previous application, the NV center used to detect the spin label is inside
a nanodiamond that is attached to the same receptor. Our results indicate that spin labels at a
distance of $10\,$nm from the NV center can be monitored with a sub-second time resolution despite
the thermal vibrations of the chemokine receptor.

The paper is organized as follows: In Sec.~\ref{Modelandstochstic}, a detailed description
of the system and the Hartmann-Hahn resonance scheme is presented. In Sec.~\ref{simulations} we
characterize the motion of the NHC-Ru complex by means of ANM simulations and introduce stochastic
differential equations for generating stochastic trajectories of the target spin. In Sec.~\ref{StochasticTreatment}
we start from the stochastic Hamiltonian of the NV-target system and derive the quantum master equation for
the coupled spins. In Sec.~\ref{detachment} we discuss the feasibility of the spin detection
considering the relaxation time of the NV spin and the NV readout photon collection efficiency.
In Sec.~\ref{CXCR4} we briefly analyze the detection of an electron spin label on a CXCR4 chemokine
receptor by using a proximate NV center in a nanodiamond. We end with the conclusions in Sec.~\ref{conclusions}.

%%%%%%%%%%%%%%%%%%%%%%%%%%%%%%%%%%%%%%%%%%%%%%%%%%%%%%%%%%%%%%%%%%%%%%
%%%%%%%%%%%%%%%%%%%%%%%%%%%%%%%%%%%%%%%%%%%%%%%%%%%%%%%%%%%%%%%%%%%%%%

\section{Model and stochastic approach}
\label{Modelandstochstic}

Starting with our main application, we consider a fluorine atom bound to the NHC-Ru
molecule, which is covalently attached to the surface of the diamond and rotates freely around its anchor
point (see Fig.~\ref{ModelSketch}). In general, the carrier molecule can be flexible, but the
NHC-Ru complex considered here exhibits only small changes in shape. The NHC-Ru molecule is immersed
in a layer of water, deposited uniformly on the surface of the diamond, and undergoes fast diffusive
motion due to thermal fluctuations (the environment being at room temperature). The NV center is implanted
close to the diamond surface and in the vicinity of (but not necessarily directly underneath) the anchor
point. We do not expect to have full control over the positions so that the details of the setup rely
on the statistical proximity between the NV center and the molecule. The axis of the NV center is assumed
to be perpendicular to the diamond surface for concrete applications in Sec.~\ref{detachment}.

%%%%%%%%%%%%%%%%%%%%%%%%%%%%%%%%%%%%%%%%%%%%%%%%%%%%%%%%%%%%%%%%%%%%%%

\subsection{Coupling between NV center and target spin}

The NV center and the fluorine atom (i.e.~the target spin) are coupled through the magnetic dipole-dipole
interaction between their electronic and nuclear spins, respectively. The NV ground state is an electronic
spin triplet (spin $S=1$) with three states having spin projections $\ket{m_s=0}$ and $\ket{m_s=\pm1}$
along the NV axis, defining the $z$-axis of our coordinate system. In what follows, the states
$\ket{m_s=0}$ and $\ket{m_s=\pm1}$ will be denoted as $\ket{0}$ and $\ket{ \pm 1}$, respectively.
The target spin is a nuclear spin doublet (spin $S=1/2$) with spin projections $\ket{\up}$ and
$\ket{\dn}$.
The degeneracy of the states $\ket{-1}$, $\ket{+1}$, both separated  from $\ket{0}$ by the zero-field splitting
$\Delta = 2.87\,\rm{GHz}$, is lifted by the external magnetic field $\bv{B}$ such that a continuous microwave
field resonant with the $\ket{0}\leftrightarrow\ket{-1}$ transition can be applied (see Fig.~\ref{ModelSketch}).
The NV center is then described within the $\ket{0}$, $\ket{-1}$ subspace by the microwave dressed states
$\ket{\pm} = (\ket{0}\pm\ket{-1})/\sqrt{2}$ separated by twice the Rabi frequency $\Omega$~\cite{cai2012robust}.

The time-varying Hamiltonian used to describe the coupled spin system in the secular
approximation and the dressed state basis is given by~\cite{cai2013diamond,london2013detecting}
\begin{equation}
\label{FullHamiltonian}
H(t) = \Omega\sigma_z^{\rm e} + \gamma_{\rm N}\bv{B}_{\rm eff}(t)\cdot\bv{\sigma}^{\rm N}
 - \sigma_x^{\rm e}[\bv{A}(t)\cdot\bv{\sigma}^{\rm N}]\,,\phantom{\int}
\end{equation}
where $\bv{A}(t)$ is the hyperfine vector, $\bv{B}_{\rm eff}(t) = \bv{B} - \frac{1}{2}\gamma_{\rm N}^{-1}\bv{A}(t)$
is the effective magnetic field and $\gamma_{\rm N}$ is the nuclear
gyromagnetic ratio. The spin operators $\sigma_j^{\rm e}$ and $\sigma_j^{\rm N}$ (with $j=x,y,z$) act on the
states $\ket{+}$, $\ket{-}$ and $\ket{\up}$, $\ket{\dn}$, respectively. The time dependence of the hyperfine
vector $\bv{A}(t)$ stems from the fluctuating position of the fluorine $\bv{r}(t)$ relative to the NV center.
Specifically, $\bv{A}(t)\equiv\bv{A}[\bv{r}(t)]$ and $\bv{r}(t)$ are related at every instant of time by
\begin{equation}\label{relationAR}
	\bv{A}(\bv{r}) = -\frac{\mu_0\gamma_{\rm e}\gamma_{\rm N}}{4\pi\vert\bv{r}\vert^3}
	(3\hat{r}_x\hat{r}_z,3\hat{r}_y\hat{r}_z,3\hat{r}_z^2-1)\,,
\end{equation}
where $\hat{r}_j$ are the components of the unit vector $\hat{\bv{r}}\equiv\bv{r}/\vert\bv{r}\vert$. Moreover,
$\gamma_{\rm e}$ and $\gamma_{\rm{N}}$ are the electronic and nuclear gyromagnetic ratios, respectively,
and $\mu_0$ is the vacuum permeability.

\begin{figure}[t]
\centering
\raisebox{2.3cm}{(a)}\hspace{5pt}\includegraphics[width=90pt]{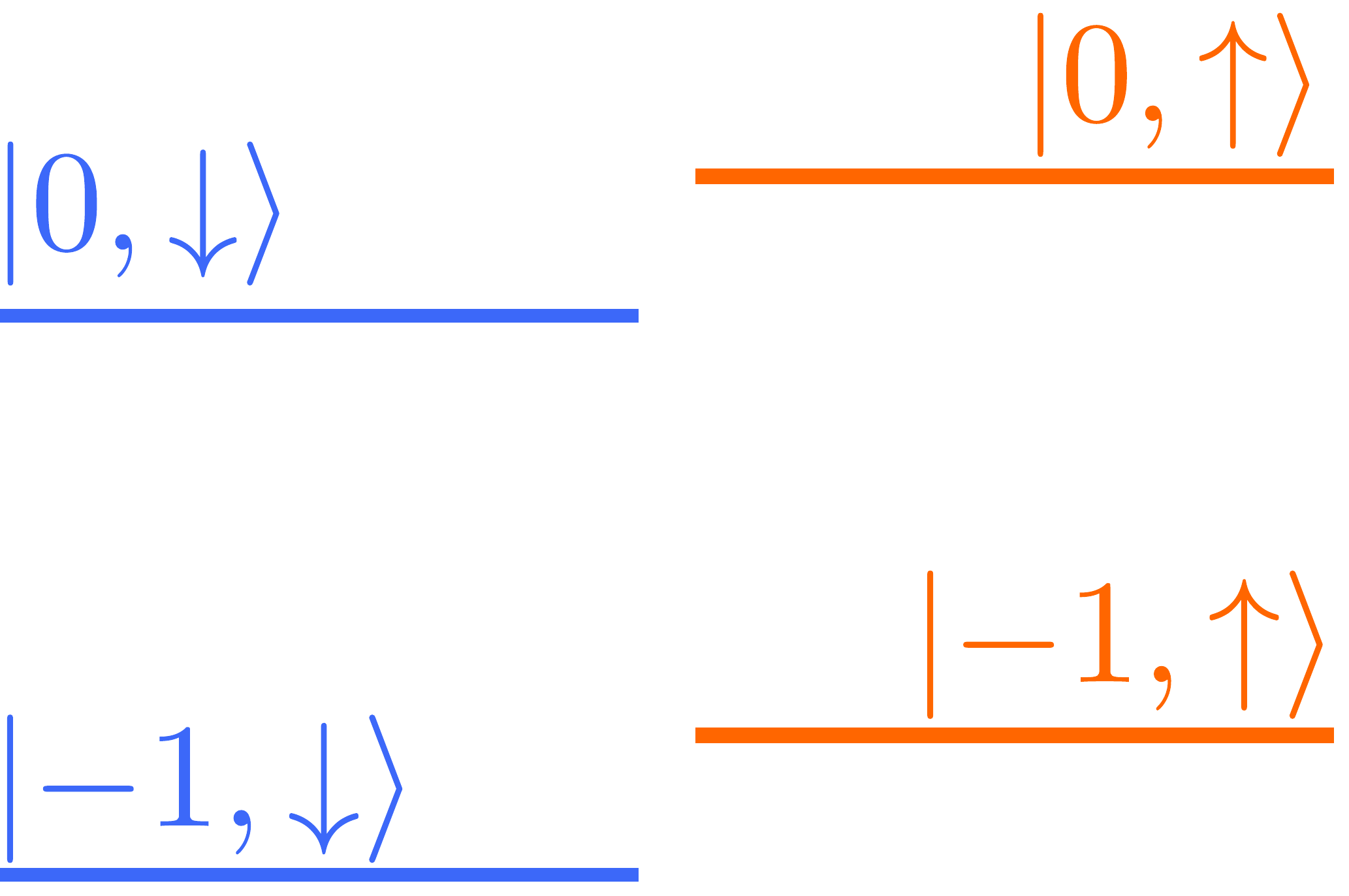}\hspace{10pt}
\raisebox{2.3cm}{(b)}\hspace{5pt}\includegraphics[width=90pt]{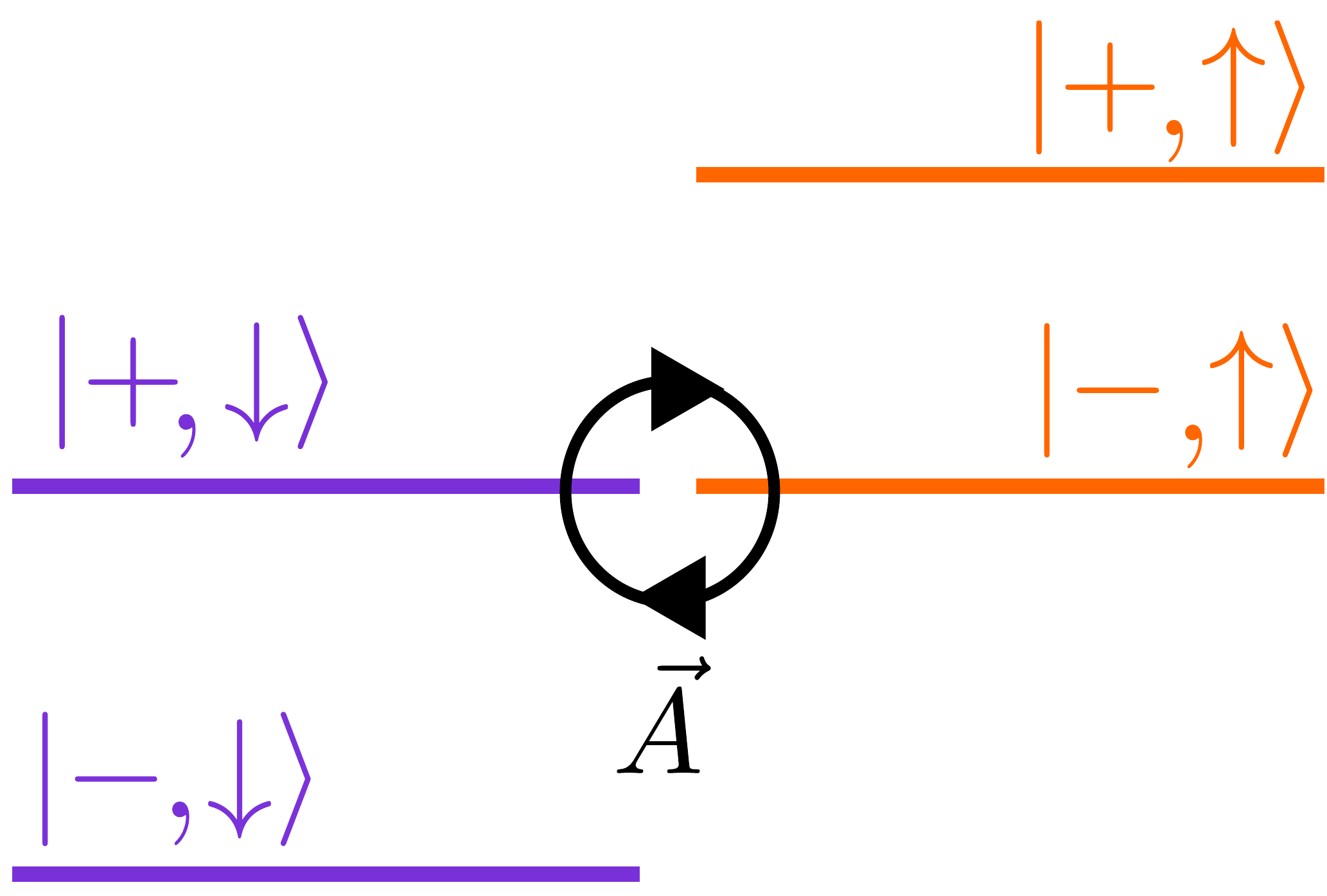}
\caption{Energy-level diagram of the electronic spin $\ket{0}, \ket{-1}$ of the NV center and the
nuclear spin $\ket{\up}, \ket{\dn}$ of the fluorine atom. (a)~The mismatch between the
electronic and nuclear energy scales suppresses the polarization transfer between the states $\ket{0,\dn}$
and $\ket{-1,\up}$ induced by the hyperfine coupling $\bv{A}$. (b)~Continuous microwave driving of the
electronic states enables polarization transfer between the dressed states $\ket{+,\dn}$ and $\ket{-,\up}$
under the Hartmann-Hahn matching condition.}
\label{hhscheme}
\end{figure}

The Hamiltonian in Eq.~\eqref{FullHamiltonian} has been shown to accurately describe the standard
HHDR scheme~\cite{hartmann1962nuclear}, which has been employed in a recent experiment to detect
\emph{static} target spins by using a single NV center~\cite{london2013detecting}. Specifically,
under the Hartman-Hahn matching condition
$\Omega-\gamma_{\rm N}\vert\bv{B}_{\rm eff}\vert=0$, the states $\ket{+,\dn}$ and $\ket{-,\up}$ of
the coupled spin system become resonant, which results in coherent polarization transfer between
them (see~Fig.~\ref{hhscheme}). In the ideal HHDR scheme, the NV spin is first optically pumped
into the state $\ket{0}$ and rotated to $\ket{+}$ with a $\pi/2$ microwave pulse. If the target spin
is in state $\ket{\dn}$ then the system is thus prepared in state $\ket{+,\dn}$ and undergoes coherent
oscillations between the $\ket{+,\dn}$ and $\ket{-,\up}$ states. The polarization transfer resulting
in the state $\ket{-,\up}$ after the interrogation time $\tau_{\rm int}$ serves as experimental indicator
for the presence of the target spin, observed as modulations in the NV fluorescence after mapping the
states $\ket{+}$, $\ket{-}$ back to $\ket{0}$, $\ket{-1}$ with a second $\pi/2$ pulse. The preparation,
polarization transfer and optical readout of the NV spin are in principle identical for static and diffusive
target spins. In practice, however, the transfer is affected by different decoherence mechanisms and the
target spin is initially in a mixed state, which will be shown to reduce the observed polarization transfer.

%%%%%%%%%%%%%%%%%%%%%%%%%%%%%%%%%%%%%%%%%%%%%%%%%%%%%%%%%%%%%%%%%%%%%%
%%%%%%%%%%%%%%%%%%%%%%%%%%%%%%%%%%%%%%%%%%%%%%%%%%%%%%%%%%%%%%%%%%%%%%

%%%%%%%%%%%%%%%%%%%%%%%%%%%%%%%%%%%%%%%%%%%%%%%%%%%%%%%%%%%%%%%%%%%%%%

\subsection{Stochastic description of diffusive target spins}

When applying the HHDR scheme to the diffusive target spin we face the additional complication that
its position $\rr(t)$ undergoes random fluctuations, i.e., Brownian motion, as a consequence of the
thermal environment. This motivates our main assumption, namely that the position $\rr(t)$ and the
hyperfine vector $\bv{A}(t)$ are stochastic processes, i.e., random functions of time. As a
consequence, the time evolution governed by $H(t)$ becomes a stochastic differential equation.

To make the problem more tractable we restrict ourselves to the case where $\rr(t)$ and $\bv{A}(t)$ are
strictly stationary stochastic processes~\cite{van1992stochastic}. This implies that $\bv{A}(t)$ has a
stationary probability distribution $p_{\bv{A}}(\sv)$ with mean $\bv{{\cal{A}}}=\avp{\bv{A}(t)}$
and variance $\bv{\sigma}^2=\avp{(\bv{A}(t)-\bv{{\cal{A}}})^2}$, where $\avp{\,\cdot\,}$ denotes
averaging with respect to $p_{\bv{A}}(\sv)$. We further assume
that $\bv{A}(t)$ is fully specified by the two-point correlations
$C_{ij}(\tau)\equiv\avt{\xi_i(t+\tau)\xi_j(t)}$ expressed in terms of the fluctuations of the hyperfine
vector $\xi_j(t) = A_j(t) - {\cal{A}}_j$, where now the average $\avt{\,\cdot\,}$ is taken over all
diffusive trajectories of the target spin. Instead of using correlations $C_{ij}(\tau)$ we can
characterize $\bv{\xi}(t)$ in an equivalent way by the power spectra
\begin{equation}
	S_{ij}(\omega) = \int_{-\infty}^{\infty}\rmd\tau\,C_{ij}(\tau)\rme^{-\rmi\omega\tau}\,.
\end{equation}
For the common case of exponentially decaying correlations
$C_{ij}(\tau)=\sigma_{ij}^2\rme^{-\vert\tau\vert/\tau_{ij}}$ the power spectra are 
$S_{ij}(\omega) = 2\sigma_{ij}^2\tau_{ij}/(1+\omega^2\tau_{ij}^2)$, where $\sigma_{ij}^2$ are the amplitudes
of the fluctuations and $\tau_{ij}$ the correlation times.

Our general approach is to divide the task of determining the evolution of the coupled spin system
into two independent parts: The first part deals with the characterization of the diffusive motion
of the target spin in terms of the distribution $p_{\bv{A}}(\sv)$ and the correlations $C_{ij}(\tau)$.
The second part consists of solving, at least approximately, the evolution of the spin system
according to $H(t)$ with the distribution $p_{\bv{A}}(\sv)$ and the power spectra $S_{ij}(\omega)$
as given ingredients.

%%%%%%%%%%%%%%%%%%%%%%%%%%%%%%%%%%%%%%%%%%%%%%%%%%%%%%%%%%%%%%%%%%%%%%
%%%%%%%%%%%%%%%%%%%%%%%%%%%%%%%%%%%%%%%%%%%%%%%%%%%%%%%%%%%%%%%%%%%%%%

\section{Diffusion of the target spin}
\label{simulations}

To efficiently characterize the diffusive motion of the molecule and the fluorine on
different time scales we use short-time ANM simulations in combination with an effective
description based on stochastic trajectories. More precisely, ANM simulations are used to
simulate the dynamics of the entire carrier molecule, whereas stochastic trajectories provide a
simple but still accurate description of the random motion of the fluorine or the target
spin in general. The time $\tau_{\rm exp}$ required for the fluorine to explore the available
spatial domain (see Fig.~\ref{trajectories}) roughly defines the time scale separating the
two descriptions.

%%%%%%%%%%%%%%%%%%%%%%%%%%%%%%%%%%%%%%%%%%%%%%%%%%%%%%%%%%%%%%%%%%%%%%

\subsection{Anisotropic network model simulations}\label{anmsim}

Numerical simulations allow us to obtain physically accurate details of the motion of the
carrier molecule to which the target spin is bound, either the NHC-Ru complex or the chemokine
receptor. Apart from ANM simulations, several alternative numerical methods have been proposed to predict
molecular motion, including molecular dynamics (MD)~\cite{alder1959md,rahman1964md} and normal mode
analysis (NMA)~\cite{cui2005normal}. ANM simulations are particularly suitable for our purposes for the
following two reasons: Considering the large disparity in time scales between the molecular motion and spin
evolution we are interested only in the slowest molecular degrees of freedom, i.e., rotations and low-frequency
vibrations, which have been shown to be accurately reproduced by the ANM. Moreover, ANM simulations are
computationally less costly than fully microscopic MD simulations and therefore applicable to
biologically relevant molecules of considerable size.

In essence, the ANM is a coarse-grained description of the collective dynamics of the molecule, independent
of the atomic details~\cite{tirion1996large,doruker2000dynamics,atilgan2001anisotropy}.
Regardless of specific atomic interactions, all atoms within a chosen cutoff distance $R_{\rm c}$ interact via
pairwise harmonic potentials whose strength is characterized by a phenomenological spring constant $\kappa$,
assumed to be the same for all atom pairs~\cite{tirion1996large}. The values of $R_{\rm c}$ and $\kappa$ are
obtained by comparing ANM results to MD simulations and experimental data~\cite{doruker2000dynamics}, where
larger cutoffs $R_{\rm c}$ (implying more interacting pairs) can be compensated by smaller spring constants
$\kappa$~\cite{atilgan2001anisotropy}. The only remaining parameters entering the simulation
are the equilibrium positions of the atoms forming the molecule, depicted in Fig.~\ref{ModelSketch} for the
NHC-Ru complex. The collisions of the molecule with the constituents of the solvent at ambient temperature
are mimicked by augmenting the ANM simulations with a Langevin dynamics, i.e., damped random kicks, for each
atom of the molecule~\cite{gillespie1996mathematics}. In this way, we have complete access to the trajectories
of the target spin, which depend on the shape and stiffness of the molecule, the solvent and its
temperature~\cite{tirado1984comparison,langer2013protein,langer2014molecular}.

When applied to the NHC-Ru molecule we use the values for the spring constant
$\kappa = 1\,{\rm kcal}/({{\rm mol\,\AA^2}})$, the cutoff distance $R_{\rm c} = 10{\rm \AA}$, the
damping coefficient of the Langevin dynamics $\zeta = 5{\rm ps}^{-1}$ and the solvent temperature
$T = 300\,{\rm K}$. The ANM simulations with this set of parameters have been tested against MD
simulations by comparing vibration spectra of various molecules and are expected to give reliable
results for the rotations of the NHC-Ru complex and the low-frequency vibrations of CXCR4 chemokine
receptor. The molecular geometry and the atomic numbers of the constituents of the NHC-Ru complex
are provided in \emph{XYZ} file format as Supplemental Material~\cite{suppmat}.

%%%%%%%%%%%%%%%%%%%%%%%%%%%%%%%%%%%%%%%%%%%%%%%%%%%%%%%%%%%%%%%%%%%%%%
\begin{figure}[t]
\centering
\raisebox{4.7cm}{(a)}\hspace{-5pt}\includegraphics[height=140pt]{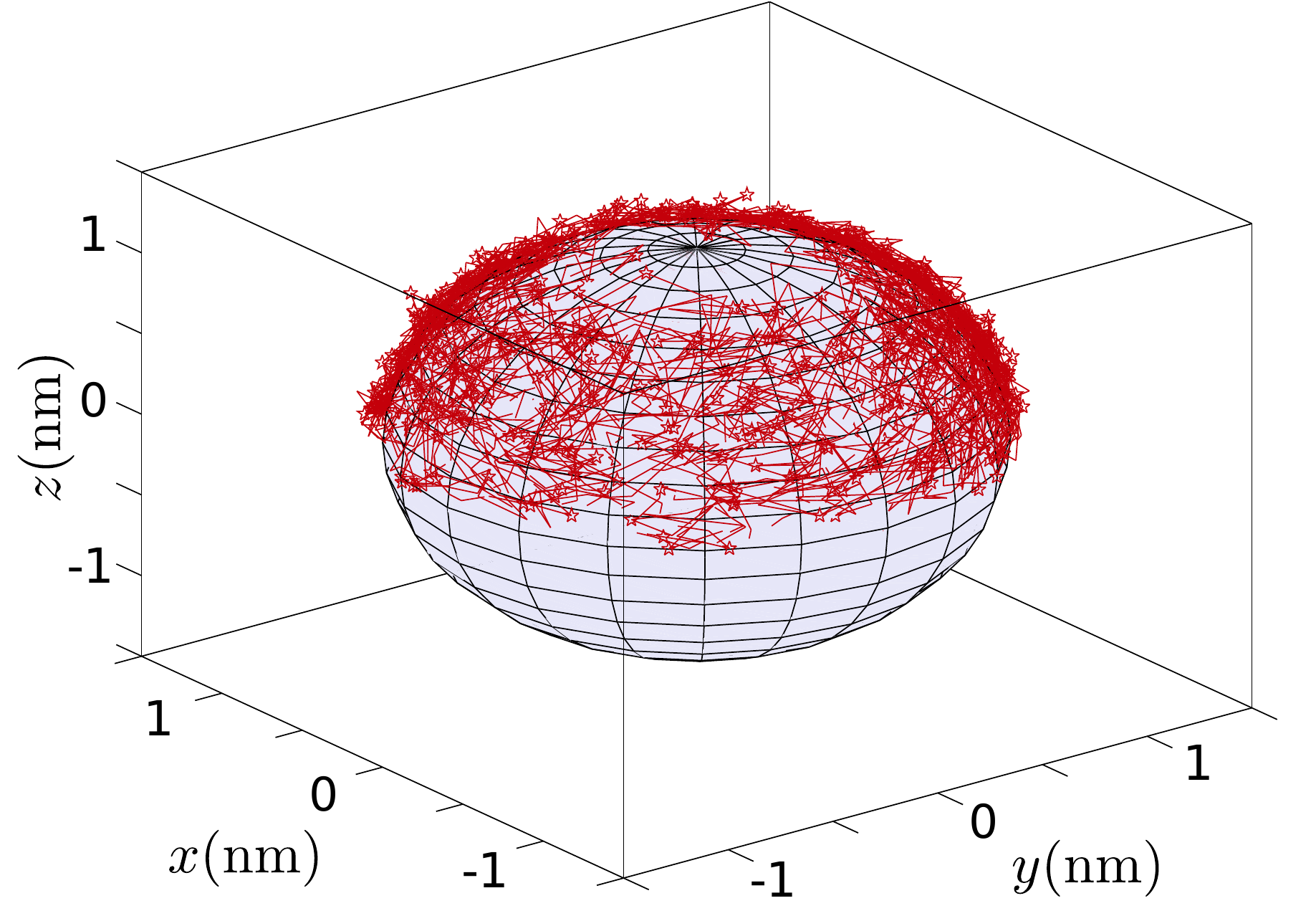}\vspace{15pt}\\
\raisebox{4.7cm}{(b)}\hspace{-5pt}\includegraphics[height=140pt]{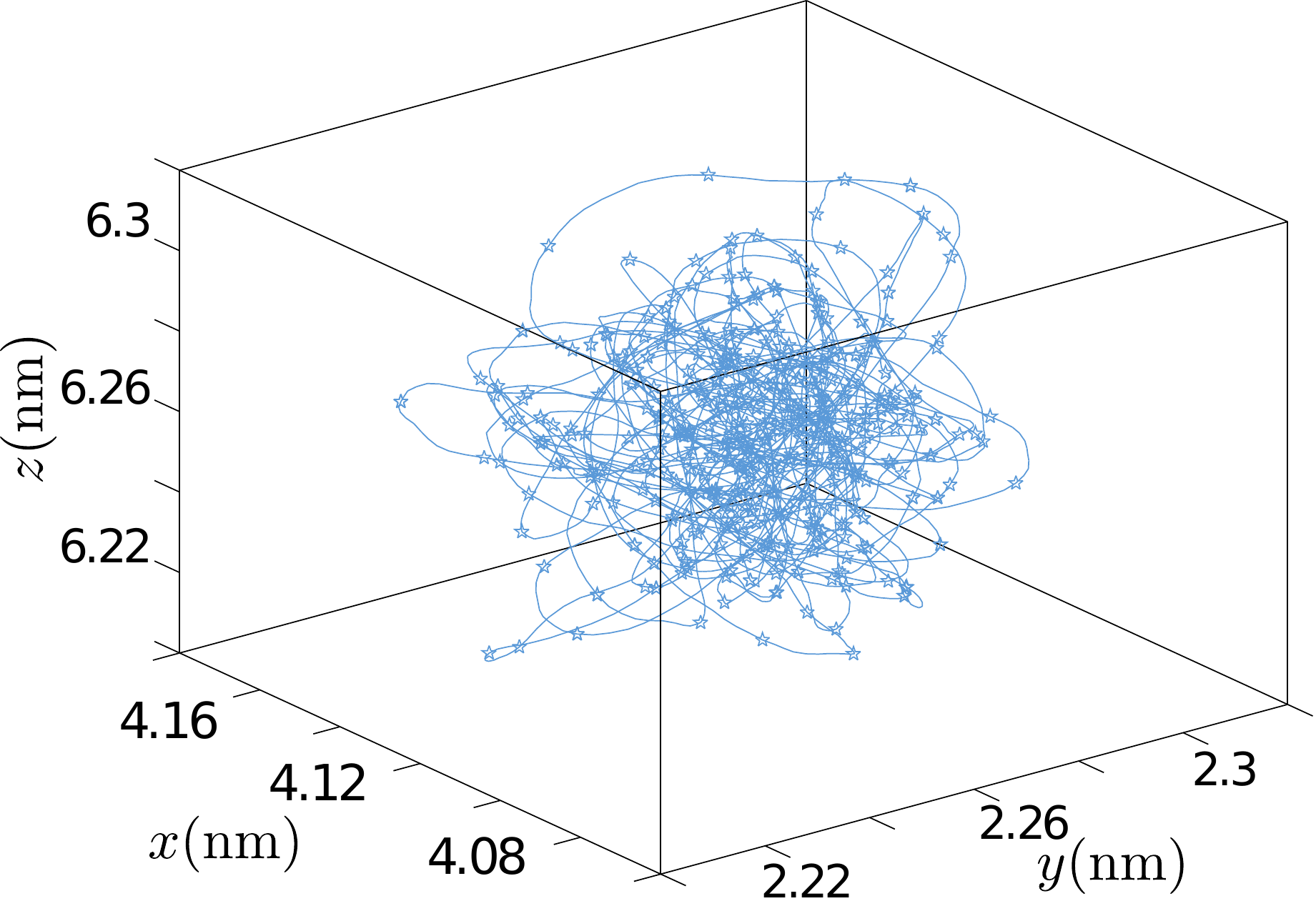}
\caption{Trajectories of the target spin obtained from short-time ANM simulations. (a)~The
fluorine atom bound to the NHC-Ru molecule follows a stochastic trajectory on a spherical
surface. (b)~The spin label attached to the chemokine receptor CXCR4 exhibits damped
fluctuations around its equilibrium position with respect to the NV center.}
\label{trajectories}
\end{figure}
%%%%%%%%%%%%%%%%%%%%%%%%%%%%%%%%%%%%%%%%%%%%%%%%%%%%%%%%%%%%%%%%%%%%%%

\begin{figure*}[t]
\centering
\raisebox{4.5cm}{(a)}\hspace{-5pt}\raisebox{0.1cm}{\includegraphics[height=120pt,width=140pt]{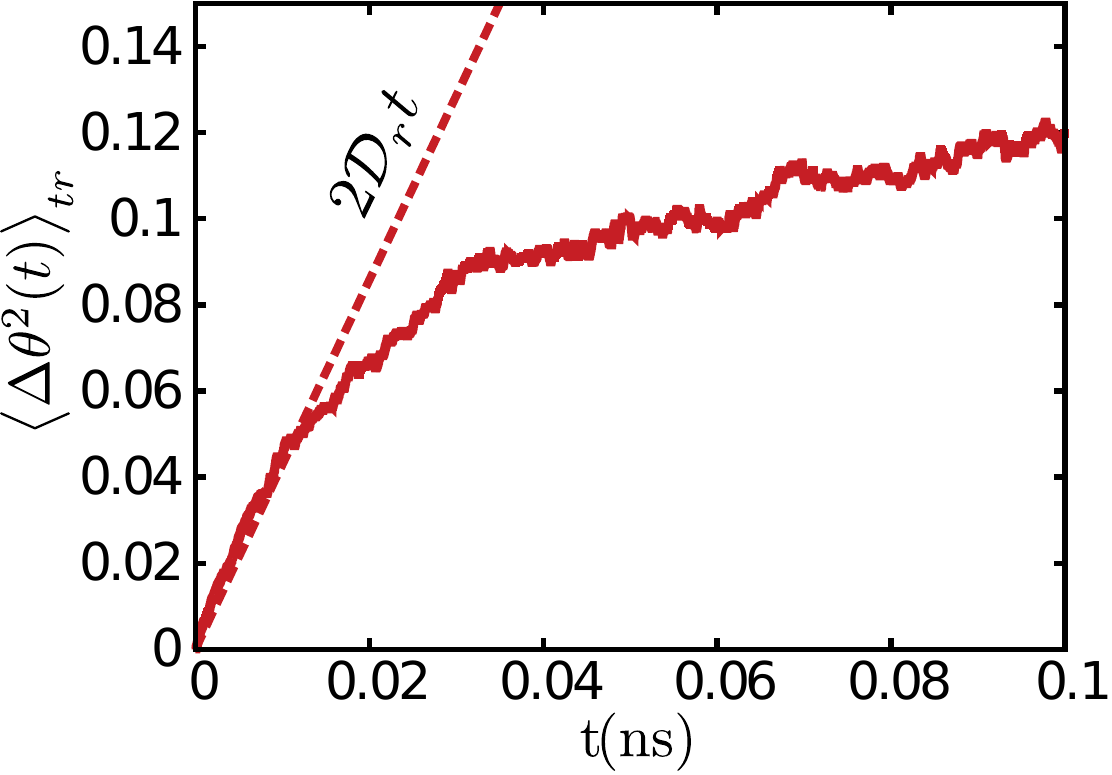}}\hspace{15pt}
\raisebox{4.5cm}{(b)}\hspace{-5pt}\includegraphics[height=125pt,width=140pt]{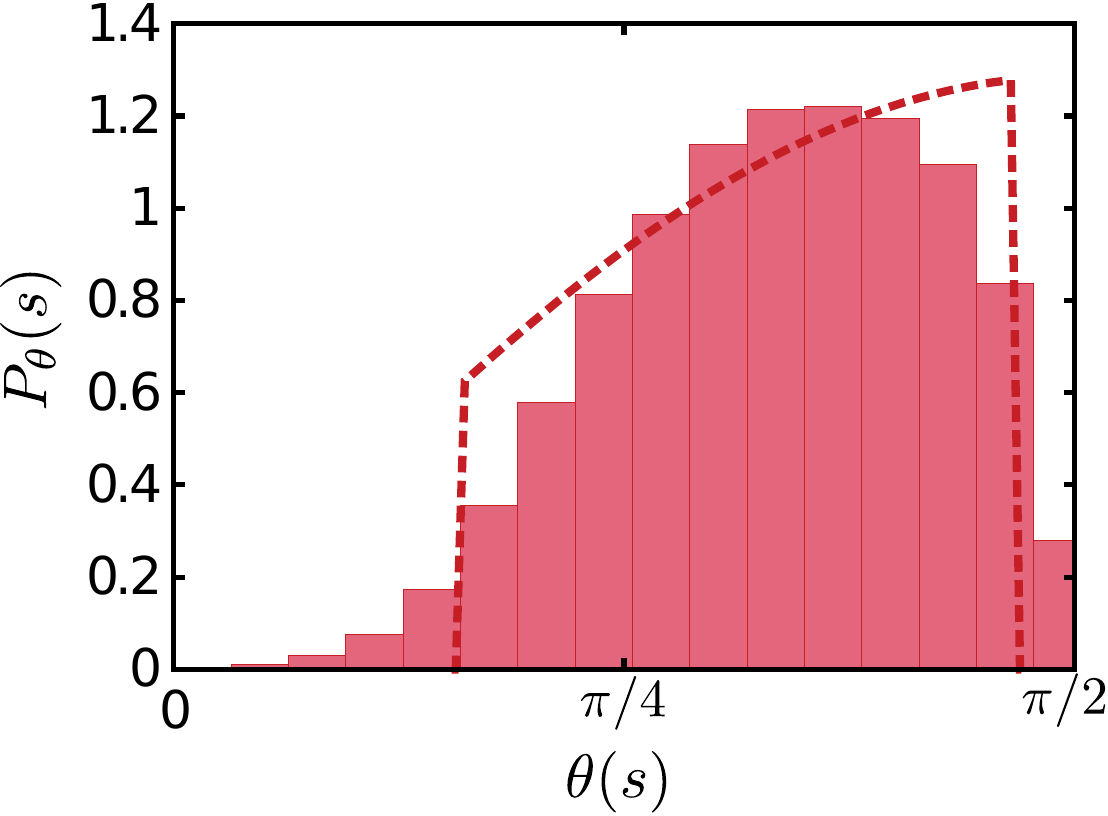}\hspace{15pt}
\raisebox{4.5cm}{(c)}\hspace{-5pt}\includegraphics[height=125pt,width=140pt]{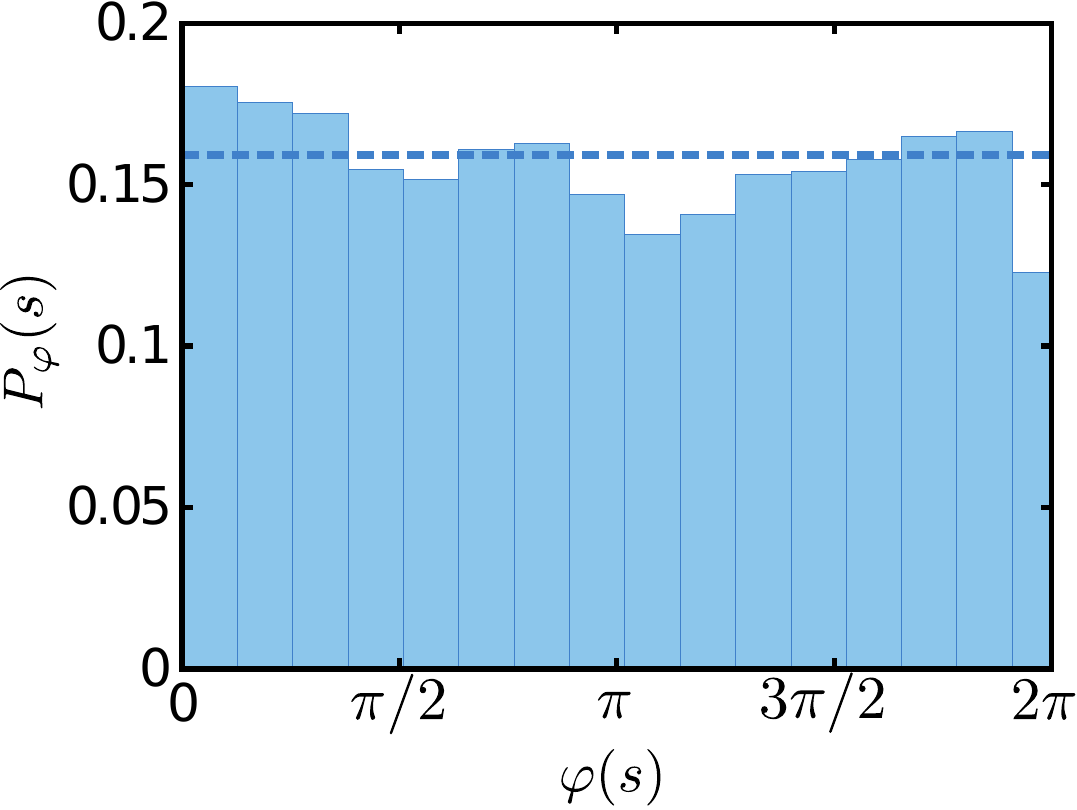}
\vspace{-1pt}
\caption{The properties of the diffusive motion of the fluorine (target spin) extracted from the
ANM simulations. (a)~The mean square displacement $\avt{ \Delta \theta^2(t)}$ as a function of
time $t$ yields the diffusion coefficient ${\cal{D}_{\rm r}}= 2.1\,{\rm ns}^{-1}$
from the short-time relation $\avt{ \Delta \theta^2(t)} = 2{\cal{D}_{\rm r}}t$. The histograms
(b) and (c) show the distributions $p_\theta(s)$ and $p_\varphi(s)$ of the polar and
azimuthal angle, respectively. The dotted lines in (b) and (c) indicate $p_{\varphi}(s)$
and $p_{\theta}(s)$ expected from a uniform distribution of the fluorine on a spherical
zone limited by the polar angles $\theta_{\rm min}$ and $\theta_{\rm max}$.}
\label{molecular}
\end{figure*}

Figure~\ref{trajectories} shows representative sample trajectories of the fluorine atom
attached to the NHC-Ru complex on the time scale of microseconds. The fluorine undergoes
a diffusive motion within a narrow spherical shell whose thickness is determined by small
oscillations of the NHC-Ru molecule. The motion is restricted to a spherical
zone as a result of the presence of the diamond surface and the spatial
extent of the molecule. If we neglect the thickness of the shell then the position of the fluorine is
conveniently described in a spherical coordinate system centered at the anchor point of the
molecule, with azimuthal angle $\varphi\in[0,2\pi]$ and polar angle $\theta\in[0, \pi/2]$.
Within this parametrization, we extract the properties of the diffusive motion in the form
of the distributions $p_\theta(s)$ and $p_\varphi(s)$ of the polar and azimuthal angle,
respectively, and the rotational diffusion coefficient $\cal{D}_{\rm r}$, as shown in
Fig.~\ref{molecular}. In particular, the mean square displacement $\avt{\Delta \theta^2(t)}$
in the regime $t\ll\tau_{\rm exp}$ allows us to determine the diffusion coefficient ${\cal{D}_{\rm r}}$
from the short-time relation $\avt{\Delta \theta^2(t)} = 2{\cal{D}_{\rm r}}t$.
The value of the diffusion coefficient extracted from ANM simulations is ${\cal{D}_{\rm r}}=2.1\,{\rm ns}^{-1}$
for the specific NHC-Ru molecule immersed in water at room temperature.

We can also estimate the diffusion coefficient ${\cal{D}_{\rm r}}$ by resorting to semi-empirical
formulas, either as a practical alternative to ANM simulations or to gain some intuition about the
molecular motion. If the approximately rod-shaped NHC-Ru molecule is modeled as a cylinder of length
$L$ and diameter $d$ and the attachment to the diamond surface is neglected then the rotational diffusion
coefficient may be obtained from~\cite{tirado1984comparison}
\begin{equation}
	{\cal{D}_{\rm r}} = \frac{3k_{\rm B} T}{\pi\eta L^3}[\log p + c(p)]\,,
\end{equation}
with $c(p) = -0.05/p^2 + 0.917/p -0.662$. The diffusion coefficient ${\cal{D}_{\rm r}}$ depends on the
ratio $p=L/d$, the Boltzmann factor $k_{\rm B} T$ and the viscosity of the solvent $\eta$. The coefficient
${\cal{D}_{\rm r}}$ depends crucially on the length $L$, implying that larger molecules undergo slower diffusion.
For a molecule of size $L=1.5$\,nm and $d=1.1$\,nm dissolved in water at $T=300\,$K with viscosity $\eta=10^{-3}\,$Pa\,s
we obtain ${\cal{D}_{\rm r}}= 0.3\,{\rm ns}^{-1}$, thereby underestimating the more accurate simulation result.

%%%%%%%%%%%%%%%%%%%%%%%%%%%%%%%%%%%%%%%%%%%%%%%%%%%%%%%%%%%%%%%%%%%%%%
%%%%%%%%%%%%%%%%%%%%%%%%%%%%%%%%%%%%%%%%%%%%%%%%%%%%%%%%%%%%%%%%%%%%%%

%%%%%%%%%%%%%%%%%%%%%%%%%%%%%%%%%%%%%%%%%%%%%%%%%%%%%%%%%%%%%%%%%%%%%%

\subsection{Stochastic trajectories and correlation functions}
\label{StochasticTrajectories}

In order to specify the correlations $C_{ij}(\tau)$ we require stochastic trajectories
on longer time scales than accessible by numerically costly ANM simulations. For that
purpose we generate stochastic trajectories of the fluorine on a spherical surface using
the stochastic differential equations in It\={o} form~\cite{brillinger2012particle}:
\begin{equation}\label{itosphere}\openup 5pt
\begin{split}
	\rmd\theta(t) &= \frac{\cal{D}_{\rm r}}{\tan\theta(t)}\,\rmd t + \sqrt{2\cal{D}_{\rm r}}\,\rmd W_1(t) \\
	\rmd\varphi(t) &= \frac{\sqrt{2\cal{D}_{\rm r}}}{\sin\theta(t)}\,\rmd W_2(t)\,,
\end{split}	
\end{equation}
where $\rmd W_1(t)$ and $\rmd W_2(t)$ are two independent Wiener processes and $\rmd t$ is the time
differential. In this setting, we neglect the spatial extent of the molecule and restrict the
motion of the fluorine to a spherical zone limited by the polar angles $\theta_{\rm min}$ and
$\theta_{\rm max}$, which is implemented through reflective boundary conditions for Eqs.~(\ref{itosphere}).
The distribution of the fluorine on the spherical zone resulting from Eqs.~(\ref{itosphere}) is uniform;
the corresponding distributions of the angles are $p_\varphi(s) = 1/2\pi$, with $s\in[0,2\pi]$, and
\begin{equation}
	p_\theta(s) = \frac{\sin(s)}{\cos(\theta_{\rm{min}}) - \cos(\theta_{\rm{max}})}\,,
\end{equation}
with $s\in[\theta_{\rm{min}},\theta_{\rm{max}}]$. The angles $\theta_{\rm min}$ and $\theta_{\rm max}$
are chosen to best fit the ANM results for $p_\theta(s)$, as illustrated in Fig.~\ref{molecular}(b).
However, the results of our analysis are rather robust with respect to the choice of $\theta_{\rm min}$
and $\theta_{\rm max}$, which alternatively may be estimated based on the shape of the molecule.

Now, for a generic stochastic process $u(t)$ evolving according to $\rmd u(t) = a[u(t)]\rmd t + b[u(t)]\rmd W(t)$,
with drift coefficient $a[u(t)]$ and diffusion coefficient $b[u(t)]$, one obtains numerically generated
trajectories by using the corresponding finite difference equation
\begin{equation}
	u(t+\delta t) = u(t) + a[u(t)]\delta t + b[u(t)](\delta t)^{1/2}N(t)\,,
\end{equation}
where $\delta t$ is the discrete time step and $N(t)$ is a temporally uncorrelated normal
random variable, i.e., $N(t)$ is statistically independent of $N(t^\prime)$.
We apply the finite difference approach to generate trajectories $[\theta(t),\varphi(t)]$ based on
Eqs.~\eqref{itosphere} with the parameters $\theta_{\rm min}$, $\theta_{\rm max}$ and $\cal{D}_{\rm r}$
obtained from ANM simulations. Trajectories of the position $\bv{r}(t)$ relative to the NV center and the
hyperfine vector $\bv{A}(t)$ are then readily found from $[\theta(t),\varphi(t)]$.

\begin{figure}[t]
\centering
\raisebox{4.8cm}{(a)}\hspace{-8pt}\includegraphics[height=130pt]{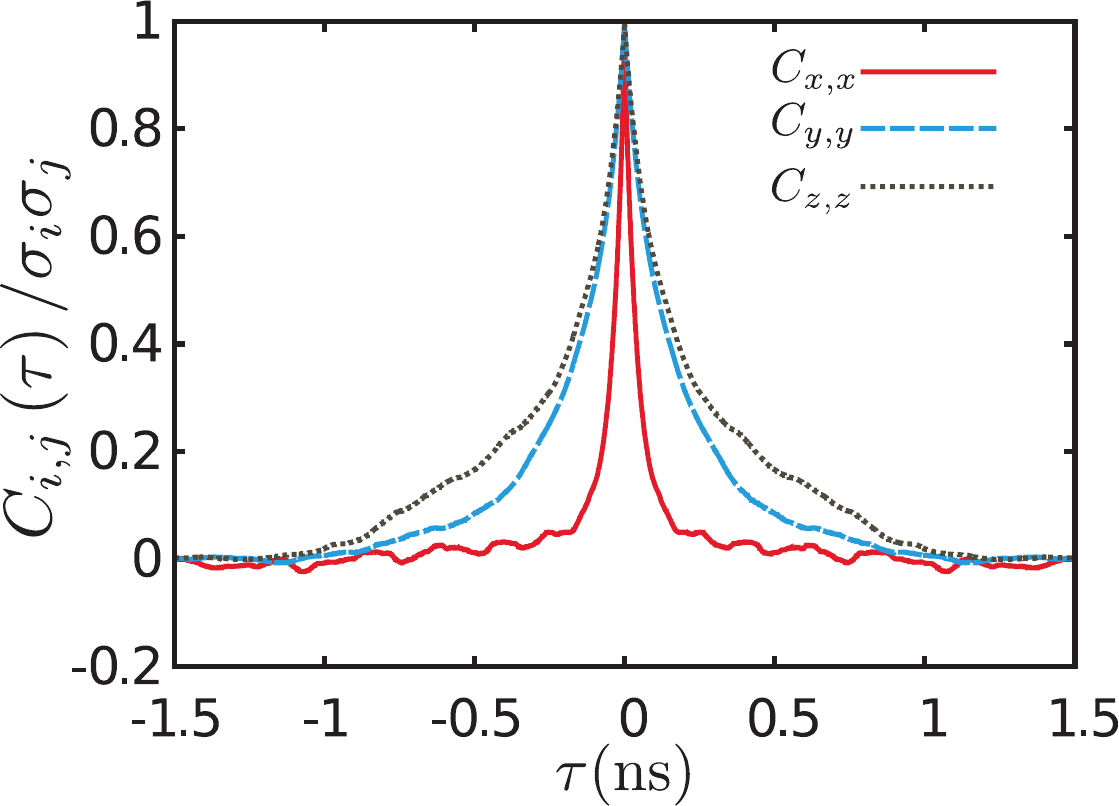}\vspace{10pt}\\
\raisebox{4.8cm}{(b)}\hspace{-8pt}\includegraphics[height=130pt]{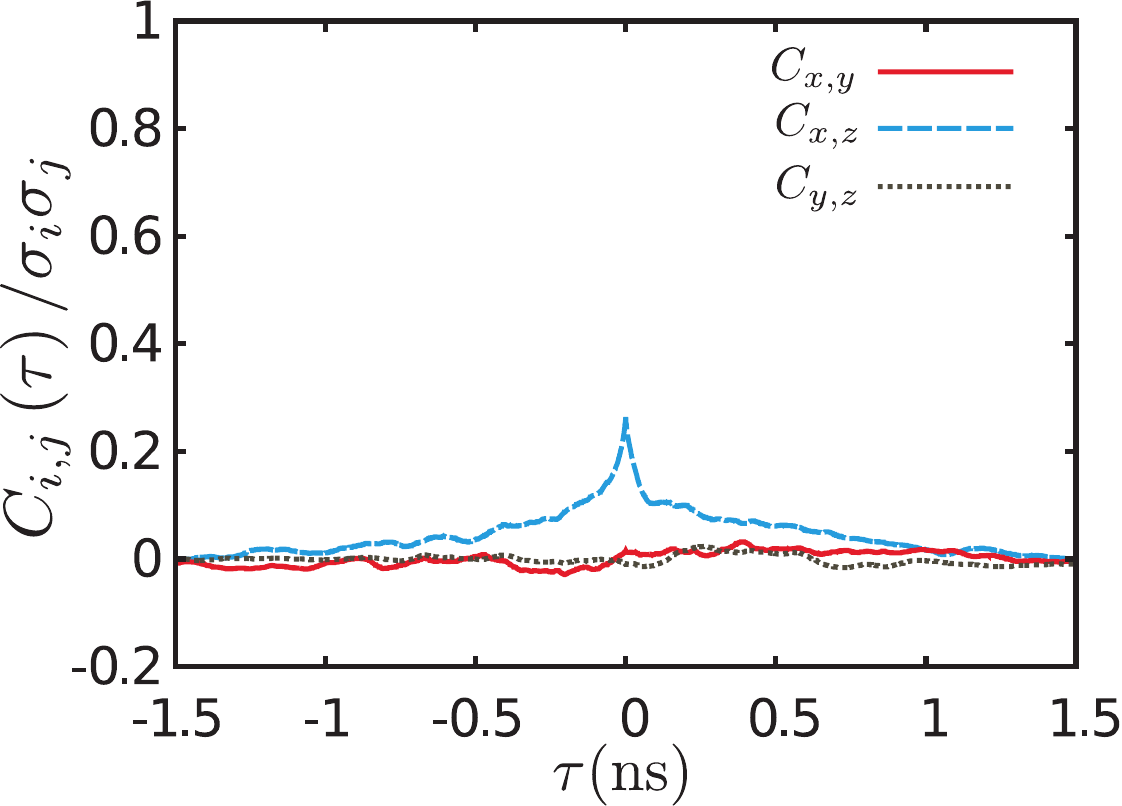}
\caption{Characterization of the thermal fluctuations $\bv{\xi}(t)$ of the hyperfine vector.
The correlations $C_{ij}(\tau)$ are normalized by the standard deviations $\sigma_{i}$
of the fluctuations. (a)~The autocorrelations $C_{ii}(\tau)$ are, to a good approximation,
exponential decay curves with a typical correlation time of $\tau_{ij}\sim 0.1\,$ns.
(b)~The cross-correlations $C_{ij}(\tau)$ with $i\neq j$ are found to be significantly
smaller than the autocorrelations.}
\label{correlations}
\end{figure}

The two-point correlations $C_{ij}(\tau)$ (with $i,j=x,y,z$) obtained from Monte Carlo sampling
over trajectories are shown in Fig.~\ref{correlations}. While the details of $C_{ij}(\tau)$
depend on the random positioning of the NV center, we find that the correlations decay quickly with
time. This ensures that the power spectra $S_{ij}(\omega)$ and integrals over the two-point correlations
$C_{ij}(\tau)$ are well defined, and that the Markov approximation is applicable in Sec.~\ref{moderate}.
In fact, the specific dependence of the stochastic equations~\eqref{itosphere} on the diffusion coefficient
$\cal{D}_{\rm r}$ implies that all correlation functions $C_{ij}(\tau)$ decay on a time scale
set by ${\cal{D}}_{\rm r}^{-1}$ and thus $\tau_{ij}\sim{\cal{D}}^{-1}_{\rm r}$ for exponentially
decaying $C_{ij}(\tau)$ (see~the~Appendix~for details).

%%%%%%%%%%%%%%%%%%%%%%%%%%%%%%%%%%%%%%%%%%%%%%%%%%%%%%%%%%%%%%%%%%%%%%
%%%%%%%%%%%%%%%%%%%%%%%%%%%%%%%%%%%%%%%%%%%%%%%%%%%%%%%%%%%%%%%%%%%%%%

\section{Evolution of the coupled spins}
\label{StochasticTreatment}

For the analysis of the stochastic evolution we split the Hamiltonian $H(t)$ into a
time-independent average Hamiltonian $H_{\cal{A}} = \avp{H(t)}$ and the remaining stochastic part
$H_\xi(t) = H(t) - \avp{H(t)}$, which read as
\begin{equation}\label{h0h1}
\begin{split}
  &H_{\cal{A}} = \Omega\sigma_z^{\rm e} + \Big(\gamma_N\bv{B} - \frac{1}{2}\bv{{\cal{A}}}\Big)\cdot\bv{\sigma}^{\rm N} - \sigma_x^{\rm e}[\bv{{\cal{A}}}\cdot\bv{\sigma}^{\rm N}] \\
  &H_\xi(t) = -\frac{1}{2}\bv{\xi}(t)\cdot\bv{\sigma}^{\rm N} - \sigma_x^{\rm e}[\bv{\xi}(t)\cdot\bv{\sigma}^{\rm N}]\,.
\end{split}
\end{equation}
We note that $H_{\cal{A}}$ is identical to the Hamiltonian for a static target spin, where the constant
hyperfine vector $\bv{A}$ is replaced by the mean $\bv{{\cal{A}}}=\avp{\bv{A}(t)}$, while $H_\xi(t)$ depends
only on the fluctuations $\bv{\xi}(t)$ of the hyperfine vector. The evolution of the density matrix $\rho(t)$
of the spin system is governed by the Liouville equation
\begin{equation}\label{Liouville}
	\frac{\partial}{\partial t}\rho(t) = [\gen_{\cal{A}} + \gen_\xi(t)]\rho(t)\,
\end{equation}
where the Liouvillian operators are defined as $\gen_{\cal{A}} = -\rmi[H_{\cal{A}},\,\cdot\;]$ and $\gen_\xi(t) = -\rmi[H_\xi(t),\,\cdot\;]$.
%%%
As it stands, Eq.~\eqref{Liouville} describes the evolution of the coupled spins for a single
diffusive trajectory of the target spin, encoded in the fluctuations $\bv{\xi}(t)$. Since we
can control neither the initial position nor the trajectories of the target spin our goal is
to determine $\avt{\rho(t)}$, i.e., the density matrix $\rho(t)$ averaged over all possible trajectories
$\bv{\xi}(t)$ with random initial conditions.

A straightforward way to calculate $\avt{\rho(t)}$ approximately is by Monte Carlo sampling: For a representative
ensemble of trajectories $\{\bv{\xi}(t)\}$ one solves Eq.~\eqref{Liouville} numerically for each $\bv{\xi}(t)$
and takes the ensemble average of the resulting density matrices to obtain $\avt{\rho(t)}$. The trajectories
$\bv{\xi}(t)$ may be generated by following the procedure outlined in Sec.~\ref{StochasticTrajectories}.
We will use Monte Carlo sampling to confirm and complement our analytical results at the end of this section.

Approximate analytical descriptions for the evolution of $\avt{\rho(t)}$ can be obtained for several relevant
parameter regimes of the system. The starting point is the stochastic Liouville equation~\cite{van1992stochastic,kubo1963stochastic}
\begin{equation}\label{VKEquation}
\begin{split}
	&\frac{\partial}{\partial t}\avt{\rho(t)} = \gen_{\cal{A}}\avt{\rho(t)}\\
	&\quad + \int_0^t\rmd\tau \avt{\gen_\xi(t)\rme^{\gen_{\cal{A}}\tau}\gen_\xi(t-\tau)}
	\rme^{-\gen_{\cal{A}}\tau}\avt{\rho(t)}\,,
\end{split}	
\end{equation}
which is obtained by formally integrating the Liouville equation~\eqref{Liouville} and performing a cumulant
expansion up to second order in $\gen_\xi(t)$. The applicability of Eq.~\eqref{VKEquation} is limited by
the amplitude of the fluctuations $\bv{\xi}(t)$ and their correlation times, which can be quantified by
the typical amplitude and width of $C_{ij}(\tau)$, denoted by $\hat{\sigma}^2$ and $\hat{\tau}$,
respectively. The cumulant expansion includes terms up to order $\hat{\sigma}^2\hat{\tau}t$,
which implies that Eq.~\eqref{VKEquation} is valid under the condition
$\hat{\sigma}^2\hat{\tau}\ll\tau_{\rm int}^{-1}$ for spin evolutions up to the
interrogation time $\tau_{\rm int}$.

We now discuss three different regimes for which analytical approximations are obtained: the 
fast-diffusion regime with $\hat{\tau}\rightarrow0$ and arbitrarily large $\hat{\sigma}^2$, the
moderate-diffusion regime with finite $\hat{\sigma}^2$ and $\hat{\tau}$, and the slow-diffusion
regime with $\hat{\tau}\gg\tau_{\rm int}$.

%%%%%%%%%%%%%%%%%%%%%%%%%%%%%%%%%%%%%%%%%%%%%%%%%%%%%%%%%%%%%%%%%%%%%%

\subsection{Fast-diffusion regime}

For vanishingly small correlation times $\hat{\tau}\rightarrow0$ the second order term in
Eq.~\eqref{VKEquation} and higher order terms of the cumulant expansion vanish. The evolution
of the spin system is governed solely by the average Hamiltonian $H_{\cal{A}}$ in Eq.~\eqref{h0h1}.
The only difference from a static target spin is that the hyperfine vector $\bv{A}$ is replaced
by the mean $\bv{{\cal{A}}}$ of the stationary process $\bv{A}(t)$ over the distribution
$p_{\bv{A}}(\sv)$. This simplifies the description of the system considerably because no
averaging over trajectories is required.

Analogously to the static case~\cite{cai2013diamond,london2013detecting} we determine the
dynamics of the spins according to $H_{\cal{A}}$ (with the magnetic field $\bv{B}$ along
the NV axis). In the vicinity of the Hartmann-Hahn matching condition, i.e., close to
vanishing \emph{averaged} detuning $\delta = \frac{1}{2}(\Omega - \gamma_{\rm N}B_z + \frac{1}{2}{\cal{A}}_z)$,
transitions to the off-resonant states $\ket{+,\up}$ and $\ket{-,\dn}$ are suppressed by the
factor $(\bv{{\cal{A}}}/\Omega)^2$ of the order $10^{-5}$. This allows us to reduce the system
to the two-dimensional subspace $\ket{+,\dn}$, $\ket{-,\up}$ with the corresponding Hamiltonian
\begin{equation}\label{fastdiff}
 	\tilde{H}_{\cal{A}} = 2 \delta\tilde{\sigma}_z - \frac{1}{2}{\cal{A}}_x\tilde{\sigma}_x
	+ \frac{1}{2}{\cal{A}}_y\tilde{\sigma}_y\,,
\end{equation}
where the Pauli matrices $\tilde{\sigma}_j$ act on the subspace only. For the nuclear spin
of the fluorine in a completely mixed state and the NV center initially in the state $\ket{+}$,
we then obtain the probability $P(t,\delta)$ of finding the NV center in the reversed state
$\ket{-}$ after the time $t$
\begin{equation}\label{avpar}
	P(t,\delta) = \frac{1}{2}\frac{\JJ^2}{\JJ^2 + \delta^2}\sin^2\big(\sqrt{\JJ^2 + \delta^2}\,t\big),
\end{equation}
where we introduced the \emph{averaged} coupling between the NV and the target spin
$\JJ=\frac{1}{4}({\cal{A}}_x^2 + {\cal{A}}_y^2)^{1/2}$. The maximal polarization
transfer occurs at resonance $\delta = 0$ after a time $t=\pi/2\JJ$, which implies
that the interrogation time $\tau_{\rm int}$ is of the order of $1/\JJ$, typically
in the millisecond regime (see also Sec.~\ref{detachment}).

%%%%%%%%%%%%%%%%%%%%%%%%%%%%%%%%%%%%%%%%%%%%%%%%%%%%%%%%%%%%%%%%%%%%%%
%%%%%%%%%%%%%%%%%%%%%%%%%%%%%%%%%%%%%%%%%%%%%%%%%%%%%%%%%%%%%%%%%%%%%%

%%%%%%%%%%%%%%%%%%%%%%%%%%%%%%%%%%%%%%%%%%%%%%%%%%%%%%%%%%%%%%%%%%%%%%

\subsection{Moderate-diffusion regime}\label{moderate}

In the moderate-diffusion regime with a finite correlation time $\hat{\tau}$, one has to take the
second order correction in Eq.~\eqref{VKEquation} into account. However, since the evolution
according to $H_{\cal{A}}$ is slow on a time scale set by $\hat{\tau}$, i.e., $\JJ\hat{\tau}\ll1$,
we can neglect the exponentials $\rme^{\pm\gen_{\cal{A}}\tau}$ and obtain
\begin{equation}\label{VKEquation2}
	\frac{\partial}{\partial t}\avt{\rho(t)} = \left(\gen_{\cal{A}} 
	+ \int_0^t\rmd\tau \avt{\gen_\xi(t)\gen_\xi(t-\tau)}\right)\avt{\rho(t)}\,.
\end{equation}
The effect of the stochastic part is best understood when Eq.~\eqref{VKEquation2} is reduced to
a Markovian master equation in Lindblad form~\cite{petruccione2002theory,rivas2012open}. In the
Markov approximation we extend the upper limit of the integral in Eq.~\eqref{VKEquation2} to
infinity and introduce the rates $\gamma_{ij} =S_{ij}(\omega = 0)$, which for the particular
case of exponentially decaying correlations $C_{ij}(\tau)=\sigma_{ij}^2\rme^{-\vert\tau\vert/\tau_{ij}}$
are given by $\gamma_{ij} = 2 \sigma_{ij}^2\tau_{ij}$. More generally, we note that the matrix
$\Gamma$ constituted by the elements $\gamma_{ij}$ is positive definite for any form of the correlations
$C_{ij}(\tau)$~\cite{cramer1940theory}. After a change to the eigenbasis of the symmetric matrix $\Gamma$,
having eigenvalues $\gamma_k$, the resulting Lindblad equation reads as
\begin{equation}\label{lindblad}
	\frac{\partial}{\partial t}\avt{\rho(t)} = -\rmi[H_{\cal{A}},\avt{\rho(t)}] + \mathcal{L}_{\rm diff}\avt{\rho(t)}
\end{equation}
with the diffusion-induced dissipative part
\begin{equation}\label{dissipator}
	\mathcal{L}_{\rm diff}\avt{\rho(t)} = \sum_{k} \frac{\gamma_{k}}{2} \Big(L_k\avt{\rho(t)}L_k^{\dagger}
	- \frac{1}{2}\{L_k^{\dagger} L_k,\avt{\rho(t)}\}\Big),
\end{equation}
where $\{\,,\:\}$ stands here for the anticommutator. The three Lindblad operators in Eq.~\eqref{dissipator}
acting on both spins are given by $L_k = L_k^{\dagger} = \breve{\sigma}_k^{\rm N}\otimes\left(\sigma_x^{\rm e}
+ \frac{1}{2}\mathbb{1}\right)$, where $\breve{\sigma}_k^{\rm N}$ are the Pauli matrices along the axes
defined by the eigenbasis of $\Gamma$. This effective description of the coupled spin system in the
form of Eqs.~\eqref{lindblad} and \eqref{dissipator} greatly simplifies the analysis of the HHDR scheme
and is applicable for a large range of the detuning $\delta$.

It can be seen from Eq.~\eqref{lindblad} that the coherent evolution is determined by the averaged
Hamiltonian $H_{\cal{A}}$, as in the fast-diffusion regime, while the diffusion-induced dissipative
part of Eq.~\eqref{lindblad} has two effects: The operators $\breve{\sigma}_k^{\rm N}$ act in three
spatial directions on the nuclear spin and result in its depolarization. The operator
$\sigma_x^{\rm e}$ acting on the NV center causes spin flips between $\ket{+}$ and $\ket{-}$, which directly
affect the Hartmann-Hahn polarization transfer. In fact, the dissipative part results from averaging
over dissimilar trajectories $\bv{\xi}(t)$, which during the interrogation time $\tau_{\rm int}$ explore
slightly different domains accessible to the hyperfine vector. This is a consequence of the finite
correlations in combination with the random initial values of $\bv{\xi}(t)$. On the other hand,
the polarization transfer when only a single diffusive trajectory of the fluorine is considered
remains always coherent.

%%%%%%%%%%%%%%%%%%%%%%%%%%%%%%%%%%%%%%%%%%%%%%%%%%%%%%%%%%%%%%%%%%%%%%
\begin{figure}[t]
\centering
\raisebox{4.53cm}{(a)}\hspace{-3pt}\includegraphics[height=130pt]{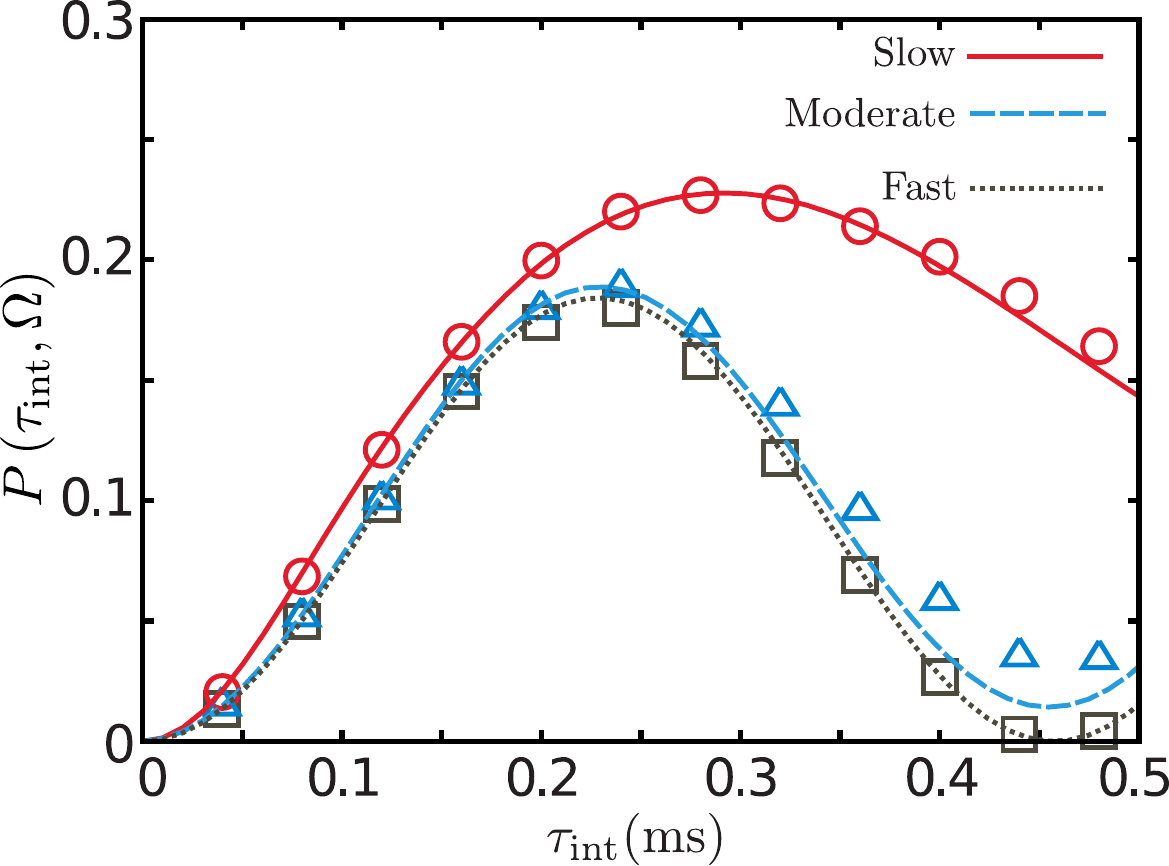}\\
\raisebox{4.53cm}{(b)}\hspace{-6pt}\includegraphics[height=130pt]{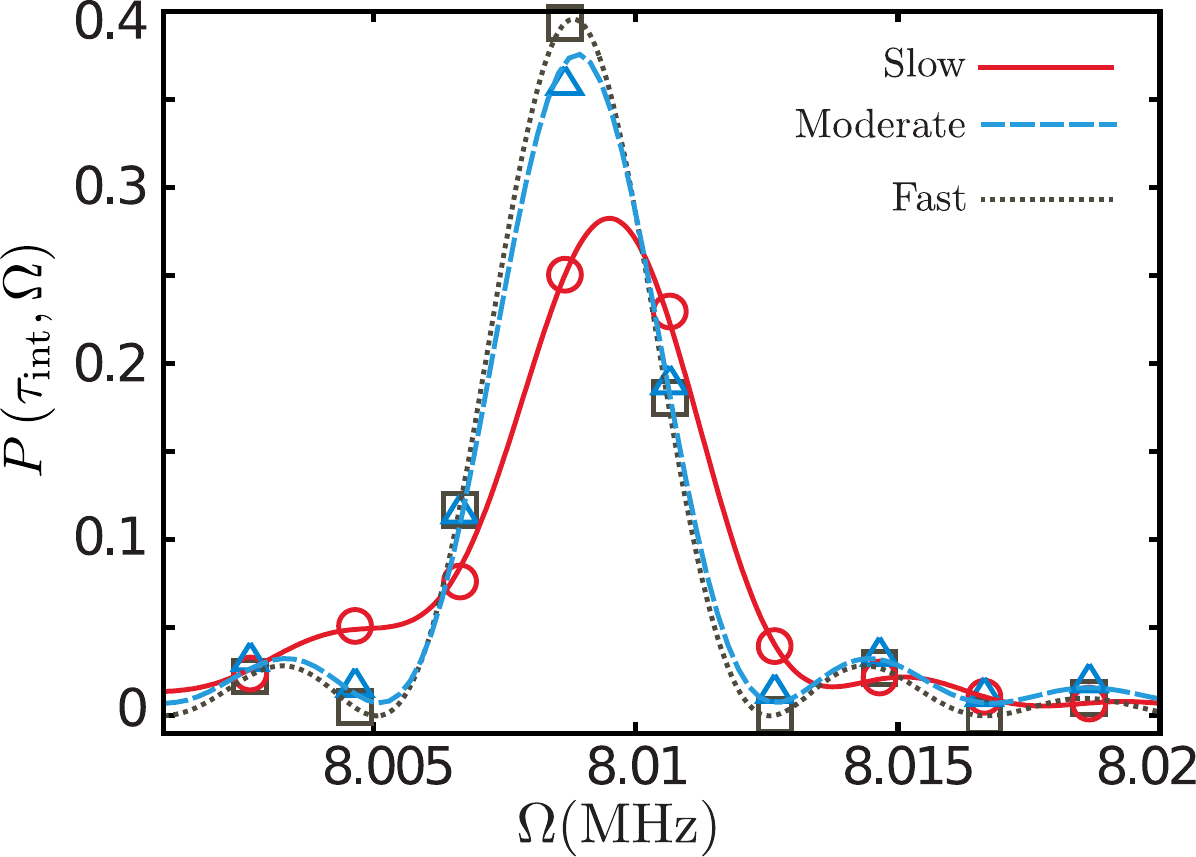}\\
\raisebox{4.53cm}{(c)}\hspace{-6pt}\includegraphics[height=130pt]{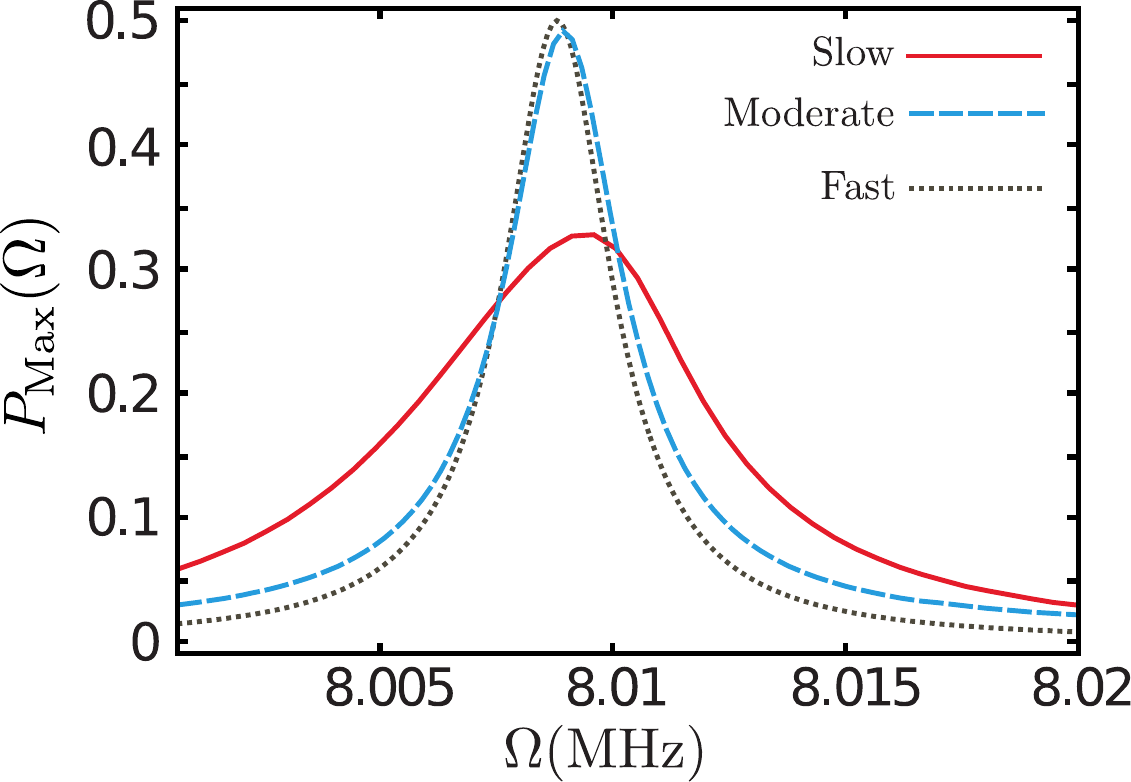}
\caption{Polarization transfer probabilities $P(\tau_{\rm int},\Omega)$ in the fast-, moderate- and
slow-diffusion regimes, obtained from Monte Carlo simulations (symbols) and analytical results (lines).
(a)~Transfer probability $P(\tau_{\rm int},\Omega)$ for fixed Rabi frequency $\Omega$ corresponding
to a detuning $\delta = 1.8\,\rm{kHz}$. Analytical results and Monte Carlo sampling over trajectories are
expected to agree in the regime $\tau_{\rm int}\ll1\,$ms. (b)~Transfer probability $P(\tau_{\rm int},\Omega)$
for a fixed interrogation time $\tau_{\rm int}=0.25\,$ms. (c)~Maximum achievable transfer probability
$P_{\rm Max}(\Omega)$ for given Rabi frequency $\Omega$ and optimized interrogation time $\tau_{\rm int}$.
Perfect polarization transfer $P_{\rm Max}(\Omega)=0.5$ is achieved for fast diffusion, while moderate and
slow diffusion result in reduced transfer. ${\cal{D}_{\rm{r}}} = (0,10,10^3)\,\rm{\mu s}^{-1}$
in (a),~(b) for the slow-, moderate- and fast-diffusion regimes, and ${\cal{D}_{\rm{r}}} = (0,1,10^3)\,\rm{\mu s}^{-1}$
in (c), with random initial positions of the target spin.}
\label{comparison}
\end{figure}
%%%%%%%%%%%%%%%%%%%%%%%%%%%%%%%%%%%%%%%%%%%%%%%%%%%%%%%%%%%%%%%%%%%%%%

%%%%%%%%%%%%%%%%%%%%%%%%%%%%%%%%%%%%%%%%%%%%%%%%%%%%%%%%%%%%%%%%%%%%%%

\subsection{Slow-diffusion regime}

In the slow-diffusion regime we assume that the hyperfine vector $\bv{A}(t)$ is constant during
the interrogation time $\tau_{\rm int}$, but drifts slowly during repeated measurements of the
Hartmann-Hahn polarization transfer. This means that $\tau_{\rm int}\ll\hat{\tau}\ll N\tau_{\rm int}$,
where $N$ is the number of individual measurement runs that are performed. This regime was also analyzed
in the context of magnetometry with rotationally diffusing nanodiamonds~\cite{maclaurin2013nanoscale}.

To obtain the averaged density matrix $\avt{\rho(t)}$ we determine the spin dynamics according
to the time-independent Hamiltonian $H$ as for the static case, but with the hyperfine
vector $\bv{A}$ as a random variable with distribution $p_{\bv{A}}(\sv)$. This yields the
density matrix $\rho(t,\bv{A})$ for each $\bv{A}$, from which we obtain
\begin{equation}\label{statav}
	\avt{\rho(t)} = \int\rmd\sv\,p_{\bv{A}}(\sv)\rho(t,\sv)\,.
\end{equation}
The density matrix $\avt{\rho(t)}$ is expected to accurately describe the observed experimental
results in the limit of large $N$.

To emphasize the difference from the fast-diffusion regime we give the explicit expression for the
probability $P(t,\delta)$ of the transition from state $\ket{+}$ to $\ket{-}$. According to the
prescription in Eq.~\eqref{statav} this probability is given by
\begin{equation}\label{avsol}
	P(t,\Omega) = \frac{1}{2}\int\rmd\sv\,p_{\bv{A}}(\sv)\frac{J^2(\sv)}{J^2(\sv) + \tilde{\delta}^2(\sv)}\sin^2
	\big[\nu(\bv{s})\,t\big],
\end{equation}
where the detuning $\tilde{\delta}(\bv{A}) = \frac{1}{2}(\Omega - \gamma_{\rm N}B_z + \frac{1}{2}A_z)$,
the coupling $J(\bv{A}) = \frac{1}{4}(A_x^2 + A_y^2)^{1/2}$ and the effective Rabi frequency
$\nu(\bv{A}) = [J^2(\bv{A}) + \tilde{\delta}^2(\bv{A})]^{1/2}$ are random variables.
In the slow-diffusion regime we thus solve the equation of motion and subsequently
average the solution over $p_{\bv{A}}(\sv)$, in contrast to the fast- and moderate-diffusion
regimes, where the equation of motion is solved for averaged detuning $\delta$ and coupling
$\mathcal{J}$.

Figure~\ref{comparison} shows the polarization transfer probabilities $P(\tau_{\rm int},\Omega)$ for
the different diffusion regimes. The general agreement between the analytical results and Monte Carlo sampling
over trajectories is illustrated in Figs.~\ref{comparison}(a) and \ref{comparison}(b) for a set of generic parameters.
The analytical results in Fig.~\ref{comparison}(a) are accurate in the regime $\tau_{\rm int}\ll1\,$ms, as
expected from the condition $\hat{\sigma}^2\hat{\tau}\ll\tau_{\rm int}^{-1}$ for the validity of the second
order cumulant expansion. The comparison in Fig.~\ref{comparison}(b) demonstrates that the analytical
approximations are valid even far from resonance. Finally, the maximum achievable transfer probability
$P_{\rm Max}(\Omega)$ for a given $\Omega$ and optimized interrogation times $\tau_{\rm int}$ is shown in
Fig.~\ref{comparison}(c). Maximal polarization transfer is achieved in the fast-diffusion regime, whereas
reduced polarization occurs in the moderate- and slow-diffusion regimes due to diffusion-induced decoherence
of the NV spin. The most detrimental effect is observed for slow diffusion, where the time scale of the
diffusion $\Dr^{-1}$ and the interrogation time $\tau_{\rm int}$ are comparable. Note that the resonance
of $P_{\rm Max}(\Omega)$ for fast diffusion (and to a good approximation for moderate diffusion) corresponds to a
Lorentzian of width $\mathcal{J}$ centered at $\delta = 0$, whereas the resonance in the slow-diffusion
regime is asymmetric and not related to the averaged detuning $\delta$.

%%%%%%%%%%%%%%%%%%%%%%%%%%%%%%%%%%%%%%%%%%%%%%%%%%%%%%%%%%%%%%%%%%%%%%
\begin{figure}[t]
\centering
\raisebox{4.9cm}{(a)}\hspace{-5pt}\includegraphics[height=135pt]{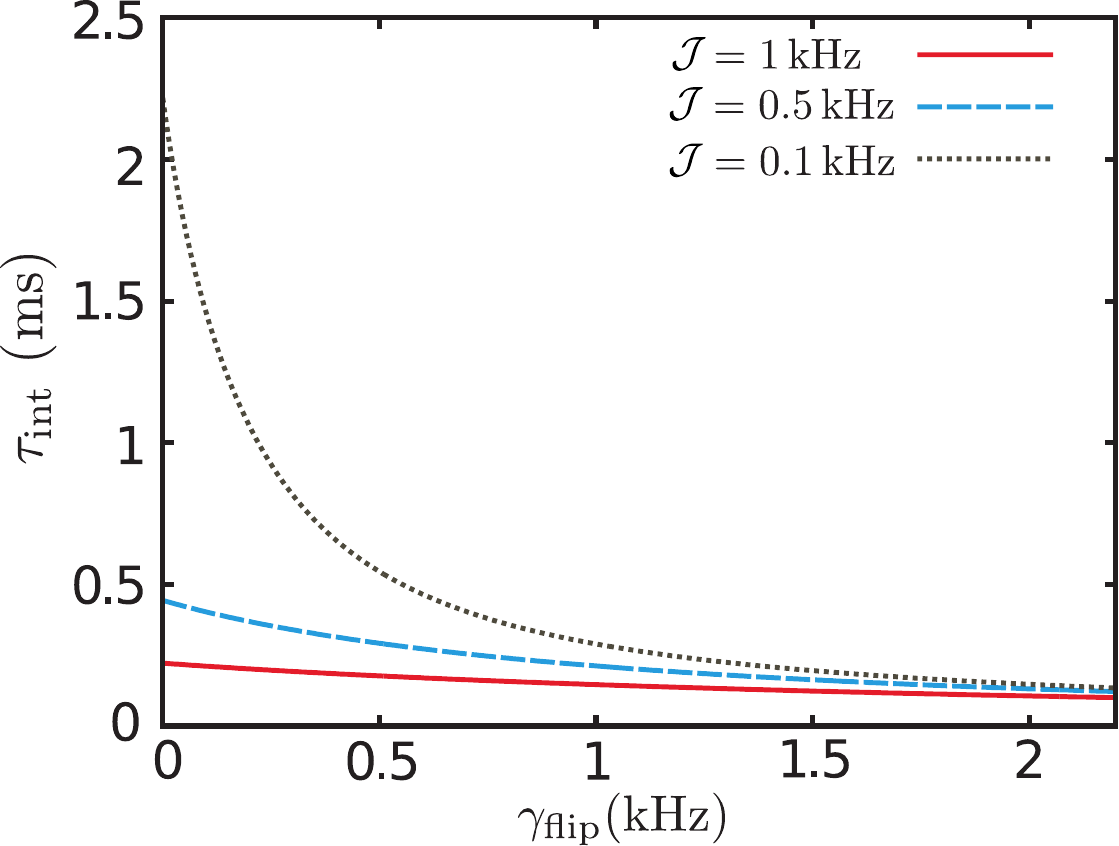}\\
\raisebox{4.8cm}{(b)}\hspace{-4pt}\includegraphics[height=131pt]{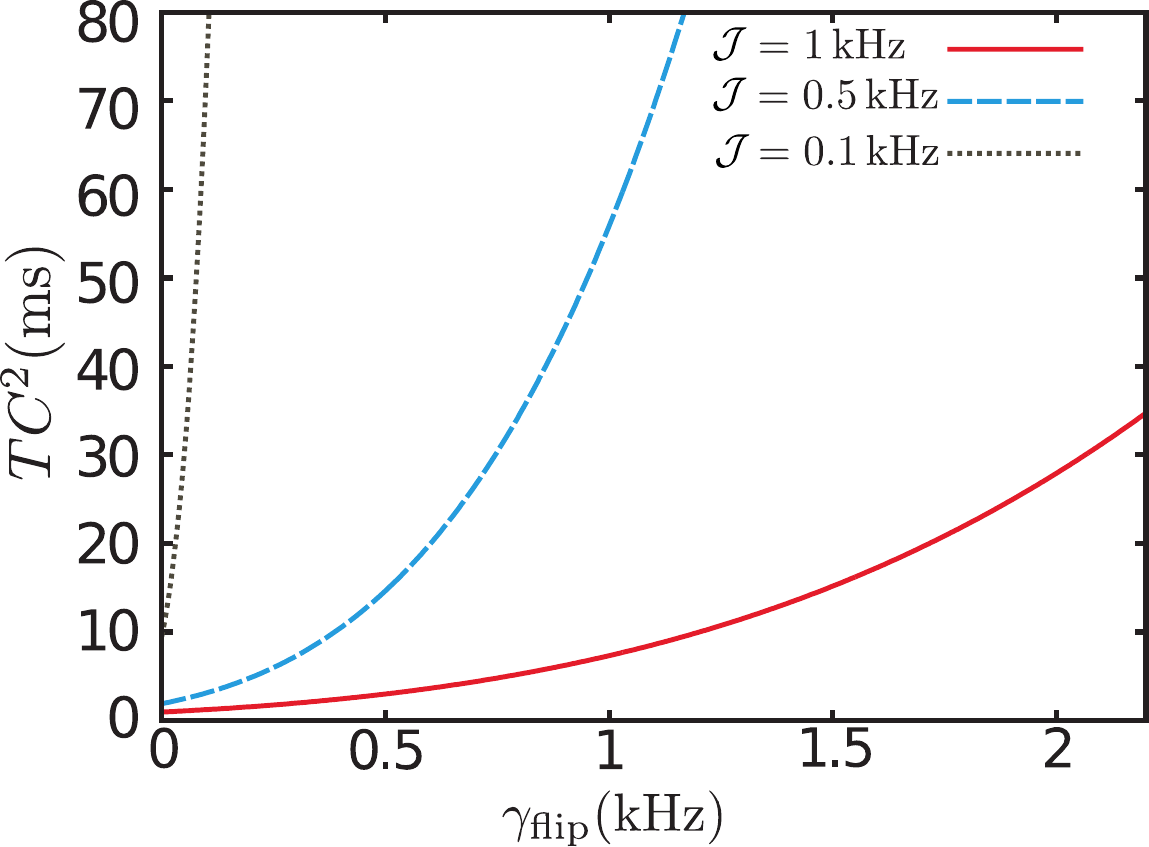}\\
\raisebox{5.0cm}{(c)}\hspace{-3pt}\includegraphics[height=140pt]{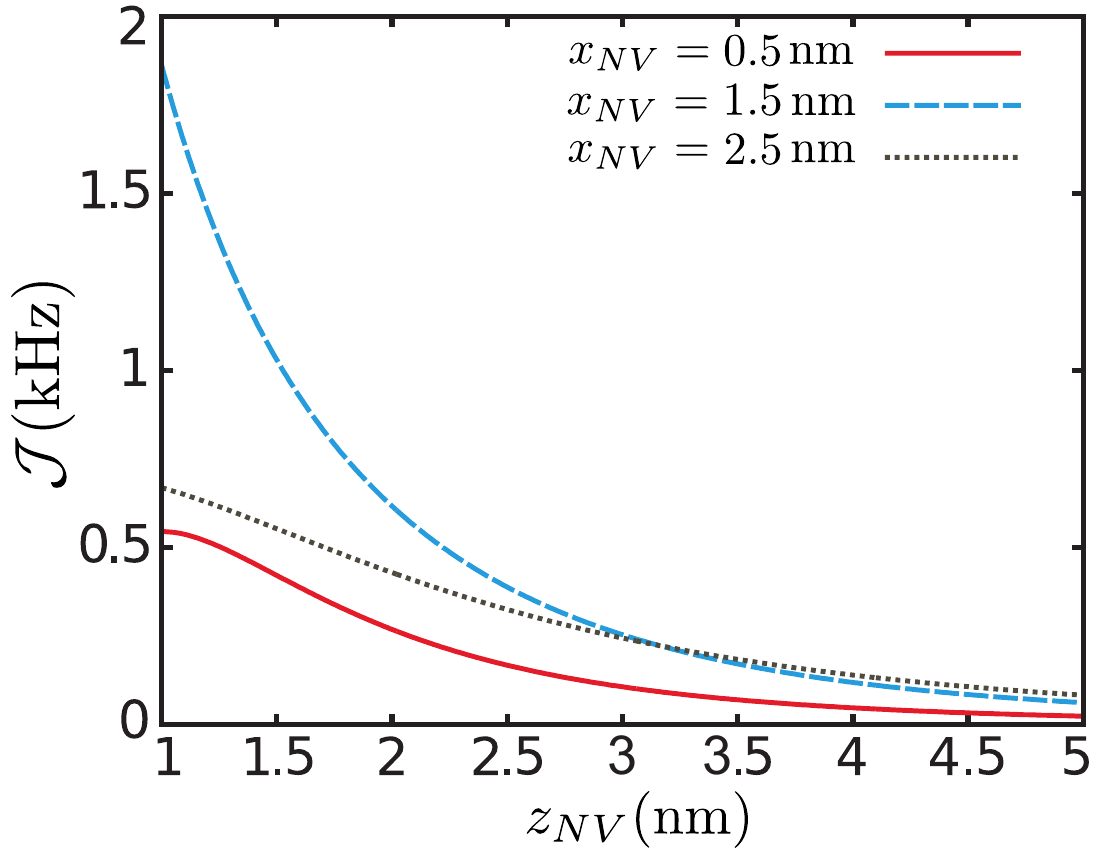}
\caption{The time-resolved detection (with unit signal-to-noise ratio) of the detachment of the fluorine
from the NHC-Ru carrier molecule. (a)~The optimal interrogation time $\tau_{\rm int}$ for each measurement
run depending on the flip rate $\gamma_{\rm{flip}}$ for different couplings $\JJ$. With increasing
$\gamma_{\rm{flip}}$ more runs with shorter interrogation times $\tau_{\rm int}$ are required. (b)~The overall
measurement time $T$ required to detect the detachment as a function of the flip rate $\gamma_{\rm{flip}}$
for different couplings $\JJ$. (c)~The dependence of averaged coupling $\JJ$ on the depth $z_{\rm NV}$
of the NV center for different lateral offsets $x_{\rm NV}$  with respect to the anchor point of the NHC-Ru
molecule (see Fig.~\ref{ModelSketch}).}
\label{signal}
\end{figure}
%%%%%%%%%%%%%%%%%%%%%%%%%%%%%%%%%%%%%%%%%%%%%%%%%%%%%%%%%%%%%%%%%%%%%%

%%%%%%%%%%%%%%%%%%%%%%%%%%%%%%%%%%%%%%%%%%%%%%%%%%%%%%%%%%%%%%%%%%%%%%
%%%%%%%%%%%%%%%%%%%%%%%%%%%%%%%%%%%%%%%%%%%%%%%%%%%%%%%%%%%%%%%%%%%%%%

\section{Detection of fluorine detachment from the molecule}
\label{detachment}

The general formalism developed in the previous section is now applied to the concrete problem
of detecting the \emph{detachment} of the fluorine atom from the carrier molecule, which demonstrates
the time-resolved operation of the HHDR scheme in a thermal environment. The situation that we
envisage here is that the fluorine-marked NHC-Ru complex serves as a catalyst for a chemical
reaction. The departure of the fluorine then indicates the event of a chemical reaction that
led to the exchange of the fluorine.

As previously shown, the diffusive motion of the fluorine attached to the molecule (in a solvent at
room temperature) exhibits short correlation times and the Hartmann-Hahn polarization transfer
therefore falls into the fast-diffusion regime, which is accurately described by the corresponding
Hamiltonian in Eq.~\eqref{fastdiff}. Nevertheless, in order to present the most general situation we
take the small diffusion-induced decoherence into account and accordingly extend the analysis to the
moderate-diffusion regime. In addition, we include incoherent processes caused by spin flips between the states $\ket{+}$
and $\ket{-}$, which are the main source of decoherence for shallowly implanted NV centers~\cite{rosskopf2014investigation}.
The effect of spin flips is described by the additional Lindblad dissipator
\begin{equation}\nonumber\label{decoherence}
	\mathcal{L}_{\rm flip}\avt{\rho(t)} = \frac{\gamma_{\rm flip}}{2}\sum_{j}\Big(L_j\avt{\rho(t)}L_j^{\dagger}
	- \frac{1}{2}\{L_j^{\dagger} L_j,\avt{\rho(t)}\}\Big),
\end{equation}
with the Lindblad operators $L_1 = \sigma_+^{\rm e}$ and $L_2 = \sigma_-^{\rm e}$,
which induce spin flips between the states $\ket{-}$ and $\ket{+}$. The flip rate
$\gamma_{\rm flip} = 1/T_{1\rho}$ is determined by the relaxation time $T_{1\rho}$
in the rotating frame~\cite{slichter1990principles}, being in the millisecond regime for
realistic experimental conditions~\cite{rosskopf2014investigation,PhysRevLett.114.017601}.
The evolution of the spin system is thus governed by the Lindblad equation
\begin{equation}\label{lindblad2}
	\frac{\partial}{\partial t}\avt{\rho(t)} = -\rmi[H_{\cal{A}},\avt{\rho(t)}] + \mathcal{L}_{\rm diff}\avt{\rho(t)}
	+ \mathcal{L}_{\rm flip}\avt{\rho(t)}\,.
\end{equation}
We note that the detrimental effect of spin flips on the polarization transfer is significantly stronger
than the diffusion-induced decoherence given that $\gamma_{\rm flip}\gg\gamma_k$. We also recall that the
initial state of the fluorine spin is completely mixed, i.e., $\rho(t=0) =  \ket{+}\!\bra{+}\otimes\frac{1}{2}\mathbb{1}$,
which reduces the observable polarization transfer by approximately a factor of two.

%%%%%%%%%%%%%%%%%%%%%%%%%%%%%%%%%%%%%%%%%%%%%%%%%%%%%%%%%%%%%%%%%%%%%%
\begin{table}[b]
\begin{center}
%\small
\renewcommand{\arraystretch}{1.3}
\renewcommand{\tabcolsep}{0.15cm}
\begin{tabular}{ccccccc}
\hline\hline
$z_{\rm NV}$ & $x_{\rm NV}$ & $\JJ$ & $C\,^*$  & ${\gamma_{\rm flip}}^{\dagger}$ &
\it $\tau_{\rm int}$ & $T$ \\
\hline

$2.0$\,nm & $1.5$\,nm & $0.7$\,kHz & $ 0.05 $ & $0.5$\,kHz & $\it 0.23${\it ms} & $\it 2.6${\it s} \\[-2pt]
$2.0$\,nm & $1.5$\,nm & $0.7$\,kHz & $ 0.05 $ & $1.0$\,kHz & $\it 0.18${\it ms} & $\it 7.9${\it s} \\[-2pt]
$2.0$\,nm & $1.5$\,nm & $0.7$\,kHz & $ 0.05 $ & $2.0$\,kHz & $\it 0.12${\it ms} & $\it 37${\it s} \\[-2pt]
$3.0$\,nm & $1.5$\,nm & $0.25$\,kHz & $ 0.05 $ & $0.5$\,kHz & $\it 0.42${\it ms} & $\it 45${\it s} \\[-2pt]
$3.0$\,nm & $1.5$\,nm & $0.25$\,kHz & $ 0.05 $ & $1.0$\,kHz & $\it 0.26${\it ms} & $\it 250${\it s} \\
\hline\hline
\multicolumn{7}{l}{\footnotesize$^*$ from Refs.~\cite{london2013detecting,taylor2008high}} \\
\multicolumn{7}{l}{\footnotesize$^\dagger$ from Ref.~\cite{rosskopf2014investigation}}
\end{tabular}
\end{center}
\vspace{-10pt}
%%%
\caption{Specific values of the model parameters and measurement times $T$ for detecting
the fluorine detachment from the NHC-Ru molecule. The coordinates $z_{\rm NV}$ and $x_{\rm NV}$ determine
respectively the depth and lateral offset of the NV with respect to the anchor point of the molecule
(see Fig.~\ref{ModelSketch}). The gyromagnetic ratio of fluorine is $\gamma_{\rm F} = 40.1\,\rm{MHz\, T^{-1}}$
and the preparation and readout time $\tau_0\sim100\,\rm{ns}$ is neglected.}
\label{TableValues}
\end{table}
%%%%%%%%%%%%%%%%%%%%%%%%%%%%%%%%%%%%%%%%%%%%%%%%%%%%%%%%%%%%%%%%%%%%%%

The time-resolved detection based on the HHDR scheme is implemented as follows:
As long as the fluorine is bound to the molecule the probability of finding the NV center
in state $\ket{-}$ after the interrogation time $\tau_{\rm int}$ is $P(\tau_{\rm int},\delta)$.
After the detachment the fluorine diffuses away very rapidly from the molecule due to its small
mass so that effectively $\JJ=0$ with the corresponding probability $P_{\JJ=0}(\tau_{\rm int},\delta)$.
The measurable signal is thus $\Delta P \equiv P \left(\tau_{\rm{int}},\delta\right)
- P_{\JJ=0} \left(\tau_{\rm{int}},\delta\right)$. The presence of the fluorine is confirmed once
the signal-to-noise ratio ($R_{\rm SN}$) exceeds unity, which for the optical readout of the NV
is equivalent to~\cite{taylor2008high} 
\begin{equation}
\Delta{P} \geq \frac{1}{C \sqrt{N}} \,,
\end{equation}
where $N$ is the total number of measurements and $C$ specifies the signal contrast and photon
collection efficiency.

The optimal time resolution is reached by tuning the interrogation time $\tau_{\rm int}$ in
dependence on the system parameters, in particular the flip rate $\gamma_{\rm flip}$. Since
the number of measurements is related to the overall measurement time $T$ through
$N = T/(\tau_{\rm int} + \tau_0)$, with $\tau_0\ll\tau_{\rm int}$ the preparation
and readout time, we obtain 
\begin{equation}
	T = \frac{\tau_{\rm int} + \tau_0}{C^2\, \Delta P^2} ,
\end{equation}
from the condition $R_{\rm SN}=1$. The main limitation to the time resolution clearly
stems from the factor $C$ and the flip rate $\gamma_{\rm flip}$ (see Table~\ref{TableValues}).

Figure~\ref{signal} shows the measurement time $T$ and the interrogation time $\tau_{\rm int}$ as
a function of the flip rate $\gamma_{\rm flip}$, as determined from Eq.~\eqref{lindblad2}. The detachment
of the fluorine can be detected (with unit signal-to-noise ratio) with a time resolution of the
order of $10\,\rm{s}$ under realistic experimental conditions. The general strategy for minimizing the
overall measurement time $T$ in the presence of spin flips at rate $\gamma_{\rm flip}$ is to
shorten the interrogation time $\tau_{\rm int}$ while increasing the number $N$ of measurement runs.
More precisely, the interrogation time $\tau_{\rm int}$ scales as $1/\gamma_{\rm flip }$ as would
be expected from a more generic analysis~\cite{huelga1997}.

It can be seen from Fig.~\ref{signal}(b) that short measurement times $T$ require sufficiently strong
couplings $\JJ$ between the NV and the fluorine spin, which in turn are determined by the depth $z_{\rm NV}$ of
the NV center and its lateral offset $x_{\rm NV}$ with respect to the NHC-Ru molecule (see Fig.~\ref{ModelSketch}).
As shown in Fig.~\ref{signal}(c), the coupling $\JJ$ decreases monotonically with the depth $z_{\rm NV}$, where
in particular an experimentally accessible depth of $z_{\rm NV}=2$\,nm corresponds to $\JJ = 0.7\,$kHz for $x_{\rm NV}=0.5\,$nm.
The dependence of $\JJ$ on the lateral offset $x_{\rm NV}$ with respect to the anchor point of the molecule is
characterized by the directionality of the dipolar spin interaction. Qualitatively speaking, the coupling $\JJ$ is
strong if the averaged position of the target spin is located within the dipolar lobes of the NV spin, whereas
$\JJ$ vanishes for the perfectly symmetric configuration $x_{\rm NV}=0$ and for large lateral offsets. 
We finally note that for a fully coherent evolution of the spins ($\gamma_{k} = \gamma_{\rm{flip}} = 0$)
the optimal interrogation time is $\tau_{\rm{int}} = \pi/2\JJ$, equivalent to a full polarization transfer.

%%%%%%%%%%%%%%%%%%%%%%%%%%%%%%%%%%%%%%%%%%%%%%%%%%%%%%%%%%%%%%%%%%%%%%
%%%%%%%%%%%%%%%%%%%%%%%%%%%%%%%%%%%%%%%%%%%%%%%%%%%%%%%%%%%%%%%%%%%%%%

\section{Probing vibrating molecules with nanodiamonds}
\label{CXCR4}

As a second application of our general approach we consider an NV center inside a
nanodiamond, which is attached to a larger molecule, here the chemokine receptor
CXCR4~\cite{wu2010structures}. The aim is to detect an electron
spin label~\cite{shi2015single} attached to the same molecule in the vicinity of the nanodiamond.
There are two main differences from the previous scenario: First, the gyromagnetic ratio of the
electron spin is three orders of magnitude larger than that of nuclear spins, which enhances the
NV-target coupling considerably. Electron spin labels up to a distance of $10\,\rm{nm}$ from the
NV can therefore be detected. Second, the relaxation time of the NV spin is shorter in nanodiamonds
than in bulk diamond because of interactions with surface spins~\cite{maclaurin2013nanoscale}.

%%%%%%%%%%%%%%%%%%%%%%%%%%%%%%%%%%%%%%%%%%%%%%%%%%%%%%%%%%%%%%%%%%%%%%
\begin{figure}[t]
\centering
\raisebox{5.0cm}{(a)}\hspace{-7pt}\includegraphics[height=136pt]{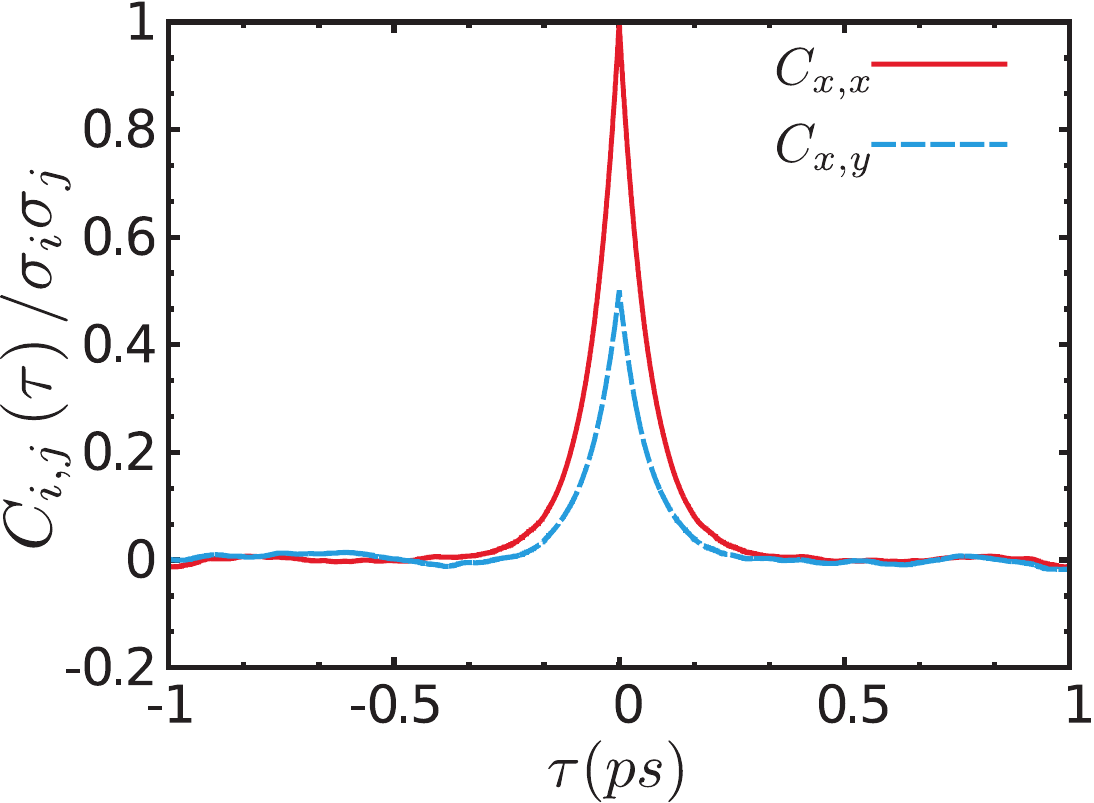}\vspace{8pt}\\
\raisebox{4.8cm}{(b)}\hspace{-7pt}\includegraphics[height=130pt]{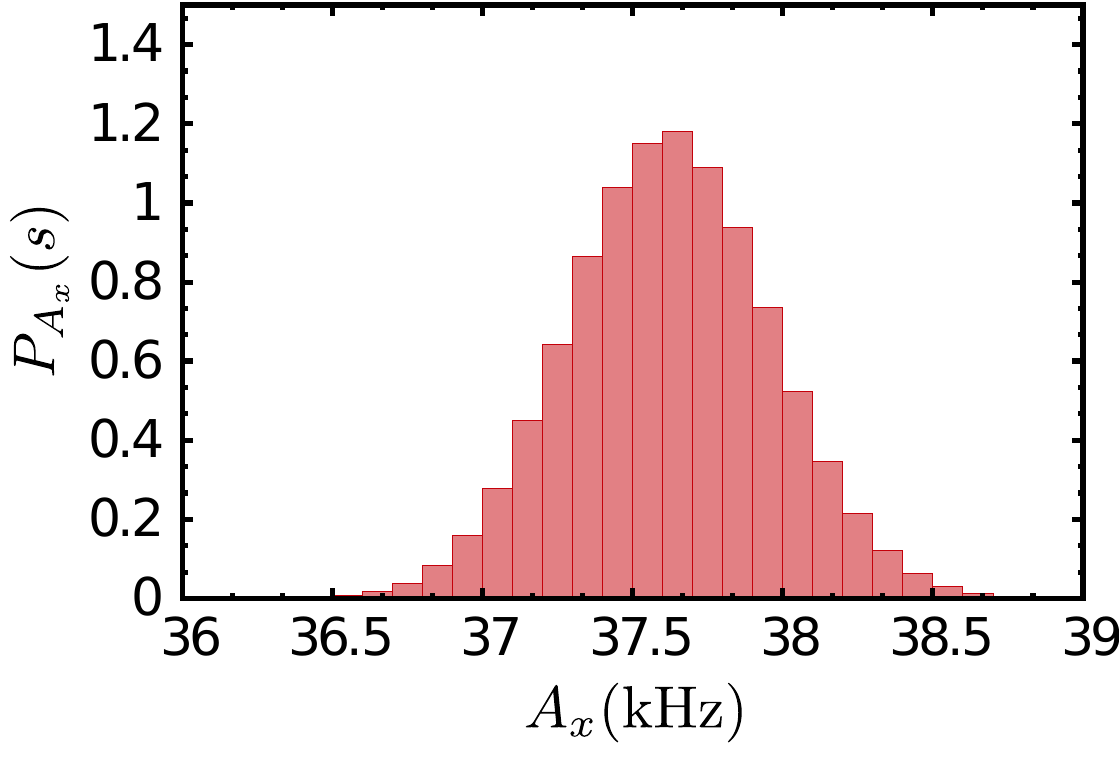}
\caption{Characterization of the thermal fluctuations $\bv{\xi}(t)$ of the hyperfine vector, assumed to
be an OU process. (a)~The correlations $C_{ij}(\tau)$ are exponentially decaying with typical
correlation times $\tau_{ij}\sim 1\,$ps. The movement of the spin label is approximately isotropic with
comparable correlations in all spatial directions. (b)~The distribution $p_{A_x}(s)$ of the $x$-component
of the hyperfine vector $\bv{A}(t)$. The approximately Gaussian shape of $p_{A_x}(s)$ is a consequence
of the underlying OU process of the fluctuations $\bv{q}(t)$.}
\label{LargeMolecule}
\end{figure}
%%%%%%%%%%%%%%%%%%%%%%%%%%%%%%%%%%%%%%%%%%%%%%%%%%%%%%%%%%%%%%%%%%%%%%

As for the previous application, we first analyze the stochastic properties of the random thermal
motion of the target spin based on ANM simulations. The parameters entering the simulation are
the same as in Sec.~\ref{anmsim} and the structural information of chemokine receptor is obtained
from the Protein Data Bank (PDB)~\cite{Berman2000}. We focus on the impact of molecular motion on
the relative distance $\bv{r}(t)$
between the NV center and the spin label under the assumption that the nanodiamond is stationary with
respect to the external magnetic field. Figure~\ref{trajectories} shows representative sample trajectories,
obtained from ANM simulations, of the relative distance $\bv{r}(t)$ between the NV center and the
spin label. The distance $\bv{r}(t)$ can be expressed as $\bv{r}(t) = \bv{R} + \bv{q}(t)$,
where $\bv{R}$ is the equilibrium position and $\bv{q}(t)$ represents the thermal fluctuations.
The numerical results suggest that $\bv{q}(t)$ can be accurately described by an Ornstein-Uhlenbeck (OU)
process with the corresponding stochastic differential equation~\cite{van1992stochastic}
\begin{equation}\label{ouprocess}
	\rmd q_k(t) = -\eta_k q_k(t)\rmd t + \sqrt{2\mathcal{D}}\,\rmd W(t),
\end{equation}
where the diffusion is assumed to be isotropic and the components $q_k(t)$ are defined with respect to
a local orthogonal basis. The restoring force of strength $\eta_k$ exerted by the molecule acts
against the thermal fluctuations which are characterized by the diffusion coefficient $\mathcal{D}$.
The stationary distribution of the fluctuations $q_k(t)$ is Gaussian with variance
$\avt{\Delta q_k^2(t)}=\mathcal{D}/\eta_k$, whereas $\avt{\Delta q_k^2(t)}=2\mathcal{D}t$ in the regime
$t\ll\tau_{\rm exp}$. We use the two relations to determine the parameters $\eta_k$ and $\mathcal{D}$
based on numerical estimates of $\avt{\Delta q_k^2(t)}$ from the simulations.
Typical values obtained in this way are $\eta_k\sim 10\,\rm{ps^{-1}}$
and $\mathcal{D}\sim 10^{-5}\,\rm{cm^2 s^{-1}}$ for an equilibrium distance $\vert\bv{R}\vert=8\,\rm{nm}$
and the environment at room temperature.

The distribution $p_{\bv{A}}(\sv)$ and the correlations $C_{ij}(\tau)$ are obtained from the stochastic
trajectories $\bv{\xi}(t)$ of the hyperfine vector, which are generated by numerically solving
Eq.~\eqref{ouprocess} with the parameters from ANM simulations and subsequently using the explicit
expression for ${\bf A}({\bf r})$ in Eq.~\eqref{relationAR}. Figure~\ref{LargeMolecule} shows the
stationary distribution $p_{A_x}(s)$ and the correlations $C_{ij}(\tau)$. The correlations decay on a time scale
$\hat{\tau}\sim 1\,\rm{ps}$ such that the condition $\hat{\sigma}^2\hat{\tau}\ll\tau_{\rm{int}}^{-1}$ for
the validity of second order cumulant expansion is easily fulfilled. We note that the correlation time is
an order of magnitude smaller than in the previous example. This is to be expected as the correlations
$C_{ij}(\tau)$ resulting from an underlying OU process decay on a time scale set by the restoring force
$\eta_k$ rather than by the diffusion coefficient $\mathcal{D}$~\cite{van1992stochastic}.

We finally proceed as in Sec.~\ref{detachment} to determine the interrogation time $\tau_{\rm int}$ and total
measurement time $T$ required to detect the spin label with unit signal-to-noise ratio. For the realistic
values of the coupling $\JJ = 20\,$kHz, the spin flip rate $\gamma_{\rm{flip}} = 1\,$kHz and $C =0.03$ we obtain
$\tau_{\rm{int}} = 0.11\,$ms and $T = 0.6\,$s. Consequently, the detection of electron spin labels attached to
the vibrating chemokine receptor in a thermal environment should be possible in experiments; with the proviso,
however, that the rotation of the nanodiamond with respect to the magnetic field is negligible~\cite{maclaurin2013nanoscale}.

%%%%%%%%%%%%%%%%%%%%%%%%%%%%%%%%%%%%%%%%%%%%%%%%%%%%%%%%%%%%%%%%%%%%%%
%%%%%%%%%%%%%%%%%%%%%%%%%%%%%%%%%%%%%%%%%%%%%%%%%%%%%%%%%%%%%%%%%%%%%%

\section{Conclusions}
\label{conclusions}

In this paper we have shown that detection of single nuclear and electronic spins undergoing diffusion
is feasible under realistic experimental conditions by using the Hartmann-Hahn double resonance scheme.
Specifically, in the case of rapidly diffusing target spins, which is particularly relevant for experiments
at room temperature, we found that the main effect of the diffusion can be absorbed in an effective
NV-target coupling. We moreover quantified the decoherence of the NV center induced by the target diffusion,
which in the limit of fast diffusion turns out to be significantly smaller than incoherent effects caused by
surface magnetic noise.

We emphasize that our approach, namely combining a statistical description of the target diffusion
with the quantum description of the spin dynamics, is very versatile. Instead of using ANM simulations
for characterizing the statistical properties of the target motion one may use more sophisticated
molecular dynamics simulations~\cite{alder1959md,rahman1964md} or, alternatively, rely on experimental
data. This is likely to be necessary for setups involving target spins in complex biological environments.

While the main focus of this paper was on spin sensing, our effective model [cf. Eqs.~(\ref{lindblad}) and
(\ref{dissipator})] is also perfectly suited for analyzing protocols for the polarization of external nuclei
by means of optically pumped NV centers~\cite{jacques2009,king2010,london2013detecting,fischer2013,chen2015optical,
rej2015hyperpolarized}. This is particularly true for modeling the dynamical polarization of gases and
liquids~\cite{abrams2014dynamic}, where the diffusive motion of the target spins plays an important role.

\vspace{-5pt}

%%%%%%%%%%%%%%%%%%%%%%%%%%%%%%%%%%%%%%%%%%%%%%%%%%%%%%%%%%%%%%%%%%%%%%
%%%%%%%%%%%%%%%%%%%%%%%%%%%%%%%%%%%%%%%%%%%%%%%%%%%%%%%%%%%%%%%%%%%%%%

\begin{acknowledgments}
The authors acknowledge discussions with F. Jelezko on experimental aspects and H. Plenio
on ruthenium based catalysts. This work was supported by an Alexander von Humboldt Professorship,
the ERC Synergy grant BioQ and the EU projects SIQS and DIADEMS.
\end{acknowledgments}

\vspace{10pt}

\appendix

\section*{Appendix: Relation between diffusion coefficient and correlation time}
\label{apprelation}

We want to clarify the relation between the correlation time of the hyperfine vector $\bv{A}(t)$
and the rotational diffusion coefficient $\Dr$. For simplicity of the argument we consider the scalar case only;
the generalization to the vectorial case is based on the same reasoning.

We assume that the hyperfine vector $A(t)$ depends on the quantity $y(t)$, corresponding to $\theta(t)$
or $\varphi(t)$ in the main text, whose evolution is described by a stochastic differential equation
of the form 
\begin{equation}\label{sdey}
	\rmd y = \mu(y)\rmd t + \sigma(y)\rmd W.
\end{equation}
The diffusion term $\sigma(y)$ depends on $\Dr$ as $\sigma(y)\sim\sqrt{\Dr}$
and we assume that the drift term $\mu(y)$ either depends on $\Dr$ as $\mu(y)\sim\Dr$
or is zero. These assumptions are compatible with evolution described by Eqs.~\eqref{itosphere}.

The stochastic differential equation for the hyperfine vector $A(t)=A[y(t)]$
is according to It\={o}'s lemma
\begin{equation}\label{sdeA}
\begin{split}
	\rmd A(t) &= \left(\mu\frac{\partial A}{\partial y} + \frac{1}{2}\sigma^2\frac{\partial^2 A}{\partial y^2}\right)\rmd t
	+ \sigma\frac{\partial A}{\partial y}\rmd W \\
	&\equiv \tilde{\mu}(A)\rmd t + \tilde{\sigma}(A)\rmd W.\phantom{\int}
\end{split}	
\end{equation}
Equation~(\ref{sdeA}) has the same form as Eq.~(\ref{sdey}) with the diffusion term
$\tilde{\sigma}(A)\sim\sqrt{\Dr}$. The drift term in Eq.~(\ref{sdeA}) necessarily
depends on $\Dr$ as $\tilde{\mu}(A)\sim\Dr$ except in the trivial case, where $\mu(y)$
is zero and $A(t)$ is simply proportional to $y(t)$.

To determine the correlations of $A(t)$ we need the probability density $p(A,t)$ governed by
the Fokker-Planck (FP) equation corresponding to Eq.~(\ref{sdeA}), which is
\begin{equation}\label{fpe}
	\frac{\partial}{\partial t}p(A,t) = -\frac{\partial}{\partial A}[\tilde{\mu}(A)p(A,t)]
	+ \frac{1}{2}\frac{\partial^2}{\partial A^2}[\tilde{\sigma}^2(A)p(A,t)]\,.
\end{equation}
The important observation is that the right-hand side of Eq.~(\ref{fpe}) is proportional
to $\Dr$ such that we can introduce the rescaled time $\tilde{t}=\Dr t$. Using the standard methods
for solving the FP equation~(\ref{fpe}) with reflecting boundary conditions one then obtains for the
autocorrelation function~\cite{gardiner1985stochastic}
\begin{equation}\label{corrdecay}
	\avt{A(\tilde{t})A(0)} = \sum_\lambda\left[\int\rmd A A P_\lambda(A)\right]^2\rme^{-\lambda\tilde{t}}\,,
\end{equation}
where $P_\lambda(A)$ are eigenfunctions with eigenvalues $\lambda$ of the time-independent FP equation.
Thus, we see from Eq.~(\ref{corrdecay}) that the correlations of $A(t)$ decay exponentially on a time
scale set by $\Dr$. We note that this result does not apply to the OU process in Eq.~(\ref{ouprocess}).

%%%%%%%%%%%%%%%%%%%%%%%%%%%%%%%%%%%%%%%%%%%%%%%%%%%%%%%%%%%%%%%%%%%%%%
%%%%%%%%%%%%%%%%%%%%%%%%%%%%%%%%%%%%%%%%%%%%%%%%%%%%%%%%%%%%%%%%%%%%%%

\section*{Supplemental Material}

The \emph{XYZ} file contains the atomic numbers and the Cartesian coordinates of the equilibrium positions of the
constituents of the NHC-Ru complex. The atom used to attach the NHC-Ru complex to the diamond is identified
as a helium atom, corresponding to a carbon atom of the diamond surface.

\vspace{-1pt}

\small

\begin{verbatim}
76

   6     0.479   -3.582    1.149  
   6    -0.564   -4.513    1.115  
   6    -1.087   -4.982   -0.094  
   6    -0.542   -4.505   -1.289  
   6     0.502   -3.575   -1.299  
   6     1.002   -3.130   -0.069  
   7     1.994   -2.229   -0.059  
   6    -2.214   -5.987   -0.116  
   6     3.317   -2.444   -0.046  
   6     3.947   -1.202   -0.022  
   6     2.861    1.787    1.146  
   6     3.011    3.178    1.131  
   6     3.442    3.856   -0.013  
   6     3.719    3.114   -1.165  
   6     3.584    1.722   -1.192  
   6     3.154    1.075   -0.025  
   6     2.387    1.087    2.398  
   6     3.854    0.957   -2.467  
   7     2.997   -0.255   -0.035  
   6     3.575    5.358   -0.028  
   6     1.824   -0.901   -0.051  
   1     3.749   -3.458   -0.045  
   1     5.022   -0.958   -0.002  
  44     0.000    0.000   -0.000  
   6     1.066   -3.066   -2.605  
   6     1.017   -3.079    2.468  
   1    -0.988   -4.886    2.061  
   1    -0.950   -4.874   -2.245  
   1    -2.796   -5.956    0.834  
   1    -2.931   -5.732   -0.931  
   1     2.776    3.756    2.039  
   1     4.041    3.646   -2.074  
   1     1.329    0.761    2.274  
   1     3.013    0.194    2.622  
   1     2.434    1.748    3.293  
   1     4.820    0.409   -2.389  
   1     3.040    0.226   -2.676  
   1     3.915    1.629   -3.353  
   1     4.651    5.643   -0.001  
   1     3.118    5.785   -0.950  
   1     3.068    5.834    0.841  
   1     0.942   -1.961   -2.679  
   1     2.150   -3.309   -2.682  
   1     0.561   -3.516   -3.490  
   1     0.868   -1.978    2.553  
   1     0.510   -3.550    3.341  
   1     2.104   -3.300    2.557  
   7    -1.670   -7.323   -0.298  
   6    -1.912   -8.060   -1.477  
   8    -1.410   -9.173   -1.631  
   2    -2.914   -7.379   -2.740  
   6    -1.749   -0.740    0.105  
   6    -2.592    0.461    0.161  
   6    -1.944    1.732    0.106  
   6    -2.736    2.880    0.306  
   6    -4.128    2.782    0.467  
   6    -4.755    1.526    0.463  
   6    -3.984    0.362    0.324  
   9    -4.920    3.985    0.651  
   8    -0.516    1.888   -0.041  
   6    -0.091    2.288   -1.366  
   6    -0.646    1.384   -2.499  
   6    -0.333    3.790   -1.662  
   1    -1.073   -7.706    0.404  
   1    -2.031   -1.743    0.164  
   1    -2.284    3.828    0.341  
   1    -5.796    1.456    0.591  
   1    -4.449   -0.580    0.359  
   1    -0.176    1.652   -3.447  
   1    -1.726    1.505   -2.594  
   1    -0.419    0.340   -2.280  
   1     0.006    4.391   -0.817  
   1    -1.390    3.982   -1.844  
   1     0.233    4.084   -2.549  
   1     0.993    2.150   -1.373  

\end{verbatim}

\bibliography{sensing_8}

\end{document}